\documentclass[nofootinbib,prl,superscriptaddress,longbibliography,a4paper,twocolumn]{revtex4-1}
\usepackage{graphicx}
\usepackage{dcolumn}
\usepackage{bm,bbm}
\usepackage{xcolor}
\usepackage{tcolorbox}
\usepackage{algorithm}
\usepackage{algpseudocode}
\usepackage{amsmath}
\usepackage{amsfonts}
\usepackage{ulem}

\usepackage{booktabs} 

\usepackage[colorlinks,breaklinks]{hyperref}
\makeatletter
\newcommand\org@hypertarget{}
\let\org@hypertarget\hypertarget
\renewcommand\hypertarget[2]{%
  \Hy@raisedlink{\org@hypertarget{#1}{}}#2%
  }

\makeatletter 
\newcommand*{\addFileDependency}[1]{ 
    \typeout{(#1)} 
    \@addtofilelist{#1} 
    \IfFileExists{#1}{}{\typeout{No file #1.}} 
} 
\makeatother 
 
\makeatother

\newcommand{\cptp}{\mathcal{E}}
\newcommand{\iden}{\mathbb{I}}

\newcommand{\tr}{\mathrm{Tr}}

\newcommand{\ccone}{\mathcal C_{\mathrm{CCTIO}}}

\newcommand{\robu}{\mathcal{R}}

\usepackage{xr}

\begin{document}
\title{Experimental Catalytic Amplification of Asymmetry}

\author{Chao Zhang}
\email{These two authors contributed equally to this work.}
\affiliation{CAS Key Laboratory of Quantum Information, University of Science and Technology of China, Hefei, 230026, China}
\affiliation{CAS Center For Excellence in Quantum Information and Quantum Physics, University of Science and Technology of China, Hefei 230026, China}

\author{Xiao-Min Hu}
\email{These two authors contributed equally to this work.}
\affiliation{CAS Key Laboratory of Quantum Information, University of Science and Technology of China, Hefei, 230026, China}
\affiliation{CAS Center For Excellence in Quantum Information and Quantum Physics, University of Science and Technology of China, Hefei 230026, China}
\affiliation{Hefei National Laboratory, University of Science and Technology of China, Hefei 230088, China}

\author{Feng Ding}
\affiliation{School of Information Science and Engineering, Shandong University, Qingdao 266237, China}

\author{Xue-Yuan Hu}
\email{xyhu@sdu.edu.cn}
\affiliation{School of Information Science and Engineering, Shandong University, Qingdao 266237, China}

\author{Yu Guo}
\affiliation{CAS Key Laboratory of Quantum Information, University of Science and Technology of China, Hefei, 230026, China}
\affiliation{CAS Center For Excellence in Quantum Information and Quantum Physics, University of Science and Technology of China, Hefei 230026, China}

\author{Bi-Heng Liu}
\email{bhliu@ustc.edu.cn}
\affiliation{CAS Key Laboratory of Quantum Information, University of Science and Technology of China, Hefei, 230026, China}
\affiliation{CAS Center For Excellence in Quantum Information and Quantum Physics, University of Science and Technology of China, Hefei 230026, China}
\affiliation{Hefei National Laboratory, University of Science and Technology of China, Hefei 230088, China}

\author{Yun-Feng Huang}
\affiliation{CAS Key Laboratory of Quantum Information, University of Science and Technology of China, Hefei, 230026, China}
\affiliation{CAS Center For Excellence in Quantum Information and Quantum Physics, University of Science and Technology of China, Hefei 230026, China}
\affiliation{Hefei National Laboratory, University of Science and Technology of China, Hefei 230088, China}

\author{Chuan-Feng Li}
\affiliation{CAS Key Laboratory of Quantum Information, University of Science and Technology of China, Hefei, 230026, China}
\affiliation{CAS Center For Excellence in Quantum Information and Quantum Physics, University of Science and Technology of China, Hefei 230026, China}
\affiliation{Hefei National Laboratory, University of Science and Technology of China, Hefei 230088, China}

\author{Guang-Can Guo}
\affiliation{CAS Key Laboratory of Quantum Information, University of Science and Technology of China, Hefei, 230026, China}
\affiliation{CAS Center For Excellence in Quantum Information and Quantum Physics, University of Science and Technology of China, Hefei 230026, China}
\affiliation{Hefei National Laboratory, University of Science and Technology of China, Hefei 230088, China}

\begin{abstract}
The manipulation and transformation of quantum resources are key parts of quantum mechanics. Among them, asymmetry is one of the most useful operational resources, which is widely used in quantum clocks, quantum metrology, and other tasks. Recent studies have shown that the asymmetry of quantum states can be significantly amplified with the assistance of correlating catalysts which are finite-dimensional auxiliaries. In the experiment, we perform translationally invariant operations, ensuring that the asymmetric resources of the entire system remain non-increasing, on a composite system composed of a catalytic system and a quantum system. The experimental results demonstrate an asymmetry amplification of $0.0172\pm0.0022$ in the system following the catalytic process. Our work showcases the potential of quantum catalytic processes and is expected to inspire further research in the field of quantum resource theories.  

\end{abstract}

\date{\today}
\maketitle

{\itshape Introduction.---}Exploring the laws of state conversion under restricted operations is a central problem in the quantum resource theory~\cite{chitambar2019quantum}. A quantum catalyst is a quantum system, which interacts with the quantum system under consideration during a given process and returns exactly to its initial state after the process~\cite{lipka2023catalysis,datta2023catalysis}. With the assistance of a catalyst, the state conversions in a variety of resource theories~\cite{lipka2023catalysis}, 
such as entanglement~\cite{PhysRevLett.83.3566} and athermality~\cite{Brandao3275CTO}, 
are greatly enriched. Since the reduced state of the catalyst remains invariant, the resource contained in the catalyst is not consumed, and the catalyst can be used in further state conversions on other quantum systems. Interestingly, it has been recently discovered that, if the catalyst is large enough and allowed to build up correlations with the system, the rate of catalytic single-copy state conversion can reach that in the asymptotic limit~\cite{PhysRevX.8.041051,PhysRevLett.126.150502, PhysRevLett.127.080502,PhysRevLett.127.150503,PhysRevLett.127.260402}. The entanglement fraction, which is the resource directly linked to the faithfulness of quantum teleportation, can even be amplified by the use of catalysts with a small dimension~\cite{PhysRevLett.127.080502}.

The quantum coherence between eigenstates of a given observable $H$ is a valuable resource of asymmetry~\cite{Marvian_2013,NC.5.3821} in the resource theory of quantum thermodynamics~\cite{Lostaglio_2019,PhysRevX.5.021001}, and finds applications in many quantum tasks such as quantum metrology~\cite{PhysRevLett.96.010401,PhysRevA.93.042107} and quantum clocks~\cite{PhysRevLett.82.2207}. The asymmetry is non-increasing under the action of translationally invariant operations (TIO)~\cite{Marvian_2013}, which are defined as completely positive and trace-preserving maps commuting with the unitary operators $e^{-iHt}$ induced by $H$. Nevertheless, the asymmetry in the system can be amplified by using finite-dimensional catalysts~\cite{PhysRevA.103.022403}, and further, when the dimension of catalysts grows to infinity, any state conversion is realizable~\cite{PhysRevLett.113.150402}. Since in reality, the systems under control are finite, it is of great interest to figure out to what extent the asymmetry can be amplified by the use of a catalyst with a small dimension, and whether this increment can be observed in experiments. This question becomes more interesting considering the recently proven undecidability in resource theories~\cite{PhysRevLett.127.270501}, namely, the general problem of characterizing the allowed transitions dictated by a set of free operations is undecidable. In this work, we experimentally observe for the first time the amplification of asymmetric resources under global TIO operation assisted by a catalytic quantum state. We also elaborate on the changes in the catalytic state and the imperfect evolution process, addressing their impact on asymmetric resources.

\begin{figure}
    \centering
    \includegraphics[width= 0.4\textwidth]{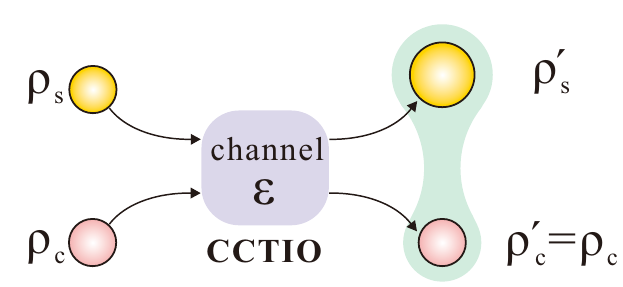}
    \caption{Schematic diagram of asymmetric resource amplification under correlating-catalytic translationally invariant operation (CCTIO). The increase of asymmetric resources is manifested as the increase of the radius of the sphere. Global TIO $\mathcal{E}$ acts on states $\rho_s$ and $\rho_c$ of system $\mathcal{S}$ and catalytic system $\mathcal{C}$. The catalytic system remains the same, that is, $\rho_c^{\prime}=\rho_c$, while the asymmetric resources of system $\mathcal{S}$ increase. Note that the final state of evolution of system $\mathcal{S}$ and auxiliary system $\mathcal{C}$ allows the correlations to be established. }
    \label{fig:enter-label}
\end{figure}

{\itshape Theory.---}Here, we consider a quantum system with Hamiltonian $H_\mathcal{S}$ and a catalyst with Hamiltoniand $H_\mathcal{C}$. Let $\rho_s$ and $\rho'_s$ be two quantum states of system $\mathcal{S}$. We say that $\rho_s$ can be transformed to $\rho'_s$ by correlating-catalytic translationally invariant operations (CCTIO) if there exists a catalyst $\rho_c$ and a global TIO $\cptp$, which is a completely positive and trance-preserving (CPTP) map satisfying $\cptp\left(e^{-i(H_\mathcal{S}+H_\mathcal{C})t}\cdot e^{i(H_\mathcal{S}+H_\mathcal{C})t}\right)=e^{-i(H_\mathcal{S}+H_\mathcal{C})t}\cptp(\cdot) e^{i(H_\mathcal{S}+H_\mathcal{C})t}$, such that
\begin{equation}
\tr_\mathcal{C}[\cptp(\rho_s\otimes\rho_c)]=\rho'_s,\ \tr_\mathcal{S}[\cptp(\rho_s\otimes\rho_c)]=\rho_c.
\end{equation}
Note that the reduced state of the catalyst is required to return exactly to its original state, while the correlation between $\mathcal{S}$ and $\mathcal{C}$ need not be erased~\cite{7404001,e19060241,PhysRevX.8.041051,PhysRevLett.122.210402,PhysRevLett.127.260402,Boes2020passingfluctuation,PhysRevLett.126.150502,PhysRevLett.127.150503,PhysRevLett.127.080502,PhysRevX.8.041016}. If $\rho'_s$ is convertible from $\rho_s$ via CCTIO, we say that $\rho'$ is in the CCTIO cone of $\rho_s$, labeled as $\rho'_s\in\ccone(\rho_s)$. Furthermore, when the dimension of $\mathcal{C}$ is restricted to $d$, the CCTIO cone of $\rho_s$ is labeled as $\ccone^{(d)}(\rho_s)$.

It has been shown that~\cite{PhysRevA.103.022403}, even when the catalyst is a qubit, the asymmetry in the state of the system $\mathcal{S}$ can still be amplified. Now we focus on the maximum amount of asymmetry in a pure qubit state that can be amplified with a qubit catalyst. To this end, we employ the (generalized) robustness $\robu$ to quantify the asymmetry~\cite{PhysRevA.93.042107}, which is defined as
\begin{equation}
    \robu(\rho)=\min_{\tau}\left\{s\geq 0\bigg|\frac{\rho+s\tau}{1+s}\in\mathcal{I}\right\},
\end{equation}
where the minimization is taken over all quantum states $\tau$, and $\mathcal{I}$ stands for the set of free states. 
When $\rho$ is a qubit state,
from~\cite{PhysRevA.93.042107}, we have $\robu(\rho)=2|\rho_{01}|$, where $\rho_{01}\equiv\langle0|\rho|1\rangle$ and $|0\rangle,|1\rangle$ are the eigenstates of $H_S$. Our problem is then formulated as
\begin{eqnarray}
\max &&\  \robu(\rho'_s)-\robu(\rho_s),\nonumber\\
\mathrm{s.t.} &&\  \rho'_s\in\ccone^{(2)}(\rho_s),\ \rho_s\in\mathcal{D}(\mathcal{H}_2).\label{eq:max_incre}
\end{eqnarray}
We solve this problem by numerical methods. Specifically, we use a bi-level programming~\cite{wen_linear_1991}. For the first level, the gradient decent~\cite{DBLP:journals/corr/Ruder16} and quasi-Newton methods~\cite{doi:10.1137/1019005} are employed to optimize over the input states $\rho_s$ and $\rho_c$. The subtask is to optimize over the global TIO $\cptp$ by semi-definite programming~\cite{doi:10.1137/1038003}. Let us show an example, in which the input state $\rho_s$, which reaches the observable increment in Eq.~(\ref{eq:max_incre}), is a pure state in the following form:
\begin{align}
    \rho_s=\frac{1}{2}(\iden+0.4333\sigma_{x}-0.9013\sigma_{z}),
\end{align}
where $\sigma_{x,y,z}$ are Pauli matrices. The corresponding catalytic state $\rho_c$ and global TIO $\cptp$ read
\begin{equation}
\rho_c=\frac{1}{2}\left(\mathbb{I}+0.5710 \sigma_{x}+0.2928\sigma_{z}\right),\label{eq:cata_state}
\end{equation}
and $\mathcal{E}(\cdot)=\mathcal{K}_{0}(\cdot) \mathcal{K}_{0}^{\dagger}+\mathcal{K}_{1}(\cdot) \mathcal{K}_{1}^{\dagger}$ with Kraus operation~\cite{doi:10.1126/science.1139892}
\begin{align}
\mathcal{K}_{0}&=\left[\begin{array}{cccc}
1 & 0 & 0 & 0 \\
0 & -0.1573 & 0.4029 & 0 \\
0 & -0.3278 & 0.8400 & 0 \\
0 & 0 & 0 & 0.7445
\end{array}\right],    \label{KrausK0}\\
\mathcal{K}_{1}&=\left[\begin{array}{cccc}
0 & \ \ \,0.9315 & 0.3636 & 0 \\
0 & 0 & 0 & 0.4721 \\
0 & 0 & 0 & 0.4721 \\
0 & 0 & 0 & 0
\end{array}\right].   \label{KrausK1}
\end{align}
It can be directly checked that the reduced state of the catalyst remains unchanged after the action of $\cptp$, while the state of the system $\mathcal{S}$ becomes
\begin{equation}
\rho^{\prime}_s=\frac{1}{2}\left(\iden+0.5314 \sigma_{x}-0.3251 \sigma_{z}\right).
\end{equation}
With the auxiliary qubit $\rho_c$ remaining unchanged, the initial quantum state $\rho_s$ of system $\mathcal{S}$ is transformed into $\rho_s’$ through global TIO $\mathcal{E}$. This transformation amplifies the quantum asymmetric resources with $\Delta \eta = \robu(\rho^{\prime}_{s})-\robu(\rho_s) = 0.0982$ of system $\mathcal{S}$, which would be impossible without catalyst $\rho_c$. This is a quantum catalytic protocol for quantum asymmetric resources based on the definition of quantum catalysis in~\cite{datta2023catalysis}.

\begin{figure*}[tph!]
\includegraphics [width= 0.9\textwidth]{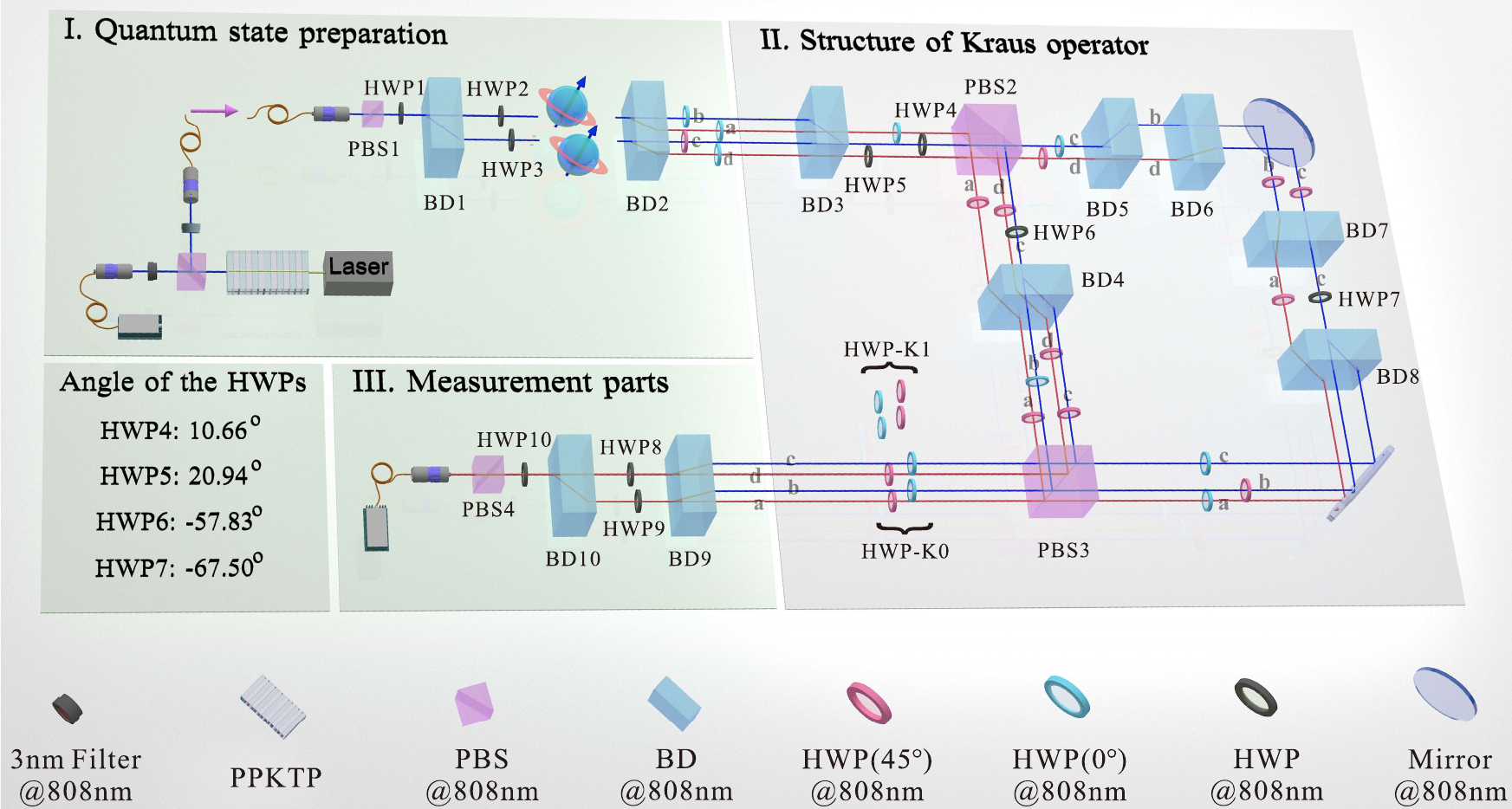}
\vspace{-0.3cm}%
\caption{Experimental setup. The blue (red) framed HWPs are set at an angle of $0^\circ$ ($45^\circ$). The black-framed HWPs require adjustment during the experiment. The QWPs are not shown but will be placed in the optical path when used. Certain beam displacers (BD1, BD3, BD4, BD5, BD8, and BD10) have optical axes in the horizontal plane, refracting horizontally polarized photons by 4~mm in the horizontal direction. Others (BD2, BD6, BD7, and BD9) have optical axes in the vertical plane, refracting vertically polarized photons by 4~mm in the vertical direction. BD: beam displacer, PBS: polarizing beam splitter, HWP: half-wave plate. 
$\mathbb{I}$: Quantum state preparation. A continuous wave laser centered at 404~nm is incident onto a 10~mm-long type-II PPKTP crystal and generates photon pairs at 808~nm. One photon is directly detected as a trigger, while the other photon is coupled into a single-mode fiber as a single-photon source. The angles of HWP1 to HWP3 are adjustable to obtain the four-dimensional quantum states required for the experiment.
$\mathbb{II}$: Structure of Kraus $\mathcal{K}_0$ and $\mathcal{K}_1$. The letters (a, b, c, d) on the ray represent the encoding of the path, where the upper layer of rays (b, c) are colored blue, and the lower layer of rays (a, d) are colored red. The quantum states reflected and transmitted by PBS3, denoted as $\rho_{\mathcal{SC}}^{\mathcal{K}_0}$ and $\rho_{\mathcal{SC}}^{\mathcal{K}_1}$, respectively, correspond to the final states resulting from the evolution of Kraus $\mathcal{K}_0$ (constructed from BD3, BD4 to PBS3) and $\mathcal{K}_1$ (constructed from BD3, BD8 to PBS3). To be consistent with the polarization on each path before BD9 with $|a_H\rangle, |b_V\rangle, |c_V\rangle, |d_H\rangle$, where the subscript $H(V)$ stands for horizontal (vertical) polarization in the path, the HWPs-group after PBS3 need to be switched to HWP-K0 (HWP-K1) to adjusts the polarization of $\rho_{\mathcal{SC}}^{\mathcal{K}_0}$  ($\rho_{\mathcal{SC}}^{\mathcal{K}_1}$). The angles of HWP4-HWP7 used to construct the channel are given in detail. $\mathbb{III}$: Measurement parts. This part carries out the tomography on the quantum states and reconstructs the density matrices of system $\mathcal{S}$ and $\mathcal{C}$.  }
\label{fig:setup}
\vspace{0.2cm}  \label{figure1}
\end{figure*}

{\itshape Protocol.---}Notably, there are two issues we have to take into account due to the unavoidable noise in experiment. 
First, the global CPTP map may deviate from TIO.
Second, the reduced state of the catalyst in the output may not return to its original state exactly. This problem arises from the catalytic state preparation and channel error in practical experiments. 

For the first issue, we evaluate the errors of asymmetry in both system $\mathcal{S}$ and catalytic system $\mathcal{C}$ caused by the deviation from TIO and subtract them from the output asymmetry in both systems. This safely excludes the effect caused by the deviation of $\mathcal{E}_{exp}$ from TIO, where $\mathcal{E}_{exp}$ is the channel in the experiment and all information of $\mathcal{E}_{exp}$ can be obtained from process topography. To tackle the second problem, we modify the input state of the catalyst, such that it is in the TIO cone of the corresponding output state~\cite{SM}. This ensures that the output state of the catalyst can return to the initial state by free operations, excluding the embezzlement phenomenon~\cite{Marvian_2013}. 

Let the input and output states of system $\mathcal{C}$ be $\rho_\mathcal{C}=\frac{1}{2}(\iden+x\sigma_x+y\sigma_y+z\sigma_z)$ and $\rho'_\mathcal{C}=\frac12(\iden+x'\sigma_x+y'\sigma_y+z'\sigma_z)$. Furthermore, the errors of the increment of asymmetry, which are caused by the deviation of the map $\mathcal{E}_{exp}$ from TIO, on systems $\mathcal{S}$ and $\mathcal{C}$, are denoted as $\epsilon_\mathcal{S}$ and $\epsilon_\mathcal{C}$, respectively. The sufficient condition can exclude the effects caused by the two noises above~\cite{SM}:
\begin{equation}
\sqrt{x^2+y^2}\max\left\{\sqrt{\frac{1-z'}{1-z}},\sqrt{\frac{1+z'}{1+z}}\right\}\leq |x'|-\epsilon_\mathcal{C}. 
\label{constraint}
\end{equation}

Furthermore, we modify the increment in the robustness of asymmetry in $\mathcal{S}$ to be $\Delta \tilde\eta \equiv \robu(\rho'_\mathcal{S})-\robu(\rho_\mathcal{S})-\epsilon_\mathcal{S}$.
This means that, if Eq.~(\ref{constraint}) is satisfied, then the increment $\Delta\tilde\eta$ observed in the experiment is caused by the catalytic transformation, instead of the noise effects.

Simultaneously, in actual experiments, the auxiliary state~(\ref{eq:cata_state}) in the example does not satisfy constraint~(\ref{constraint}) after evolution. To match the quantum evolution $\mathcal{E}_{exp}$ prepared in actual experiments, we modify the initial state of the catalyst (detailed in~\cite{SM}) to be
\begin{equation}
    \rho_{\mathcal{C}}=\frac{1}{2}\left(\mathbb{I}+0.4410 \sigma_{x}+0.2928 \sigma_{z}\right).\label{fuzhustate}
\end{equation}
Thus, the initial catalyst state $\rho_\mathcal{C}$ is prepared as a mixed state~\cite{PhysRevLett.127.220501} $\rho_{\mathcal{C}}=c_0\cdot\rho_{c_0}+c_1\cdot\rho_{c_1}$, with $c_0=0.7306, c_1=0.2694$ and 
\begin{equation}
    \rho_{c_0}=|\psi_{c_0}\rangle \langle\psi_{c_0}|,\quad\quad \rho_{c_1}=|\psi_{c_1}\rangle\langle\psi_{c_1}|,     \notag
\end{equation}
where 
\begin{equation}
    |\psi_{c_0}\rangle=(0.8040,0.5946)^{\dagger}, \quad|\psi_{c_1}\rangle=(0.8040,-0.5946)^{\dagger}.  \notag
\end{equation}

Additionally, the state $\rho_S=\rho_s$ of system $\mathcal{S}$ remains unchanged and is a pure state. Thus, the state $\rho_\mathcal{S} \otimes \rho_\mathcal{C}$ is decomposed into 
\begin{equation}
    \rho_\mathcal{S} \otimes \rho_\mathcal{C}=0.7306 \cdot \rho_{\mathcal{S}}\otimes \rho_{c_0}+0.2694 \cdot \rho_{\mathcal{S}}\otimes \rho_{c_1}.  \label{xishu}
\end{equation}

Through the evolution of $\mathcal{E}$, the equation
\begin{equation}
    \mathcal{E}(\rho_\mathcal{S}\otimes\rho_\mathcal{C})=\mathcal{K}_{0}(\rho_\mathcal{S}\otimes\rho_\mathcal{C})\mathcal{K}_{0}^{\dagger}+\mathcal{K}_{1}(\rho_\mathcal{S}\otimes\rho_\mathcal{C})\mathcal{K}_{1}^{\dagger}
\end{equation} 
can be decomposed into
\begin{align}
    \mathcal{E}(\rho_\mathcal{S}\otimes\rho_\mathcal{C})=\\ \notag
   0.7306\mathcal{K}_{0}(\rho_{\mathcal{S}}&\otimes\rho_{c_0})\mathcal{K}_{0}^{\dagger}+0.7306\mathcal{K}_{1}(\rho_{\mathcal{S}}\otimes\rho_{c_0})\mathcal{K}_{1}^{\dagger}+\\\notag
    0.2694\mathcal{K}_{0}(\rho_{\mathcal{S}}&\otimes\rho_{c_1})\mathcal{K}_{0}^{\dagger}+0.2694\mathcal{K}_{1}(\rho_{\mathcal{S}}\otimes\rho_{c_1})\mathcal{K}_{1}^{\dagger}.
\end{align}

In experiment, we construct an unabridged evolution $\mathcal{E}$ for the input state $\rho_{\mathcal{S}}\otimes\rho_{c_0}$ and $\rho_{\mathcal{S}}\otimes\rho_{c_1}$. Based on the coefficients in (\ref{xishu}), we construct a classical mixture of probabilities over time with two terms.

{\itshape Experiment.---} We choose systems $\mathcal{S}$ and $\mathcal{C}$ to be both two-dimensional and implement the CCTIO evolution in the experiment. We divide the experiment into three steps:  1. Characterize the process matrix $\mathbf{M}_p$ of evolution channel $\mathcal{E}$ from process tomography~\cite{PhysRevA.68.012305} and determine the catalytic state according to specific $\mathbf{M}_p$. 2. Prepare and measure the initial quantum states $\rho_{\mathcal{S}}$ and $\rho_{\mathcal{C}}$. 3. Perform the $\mathbf{M}_p$ operation, and carry out quantum state tomography~\cite{PhysRevA.66.012303} on the output quantum states. 

As shown in Fig.~\ref{figure1}.
Part I: State preparation.
The photon pairs are generated by the type-II spontaneous parametric down-conversion process. One photon is detected as a trigger, and the other photon enters our experimental setup. The state $\rho_{\mathcal{S}}\otimes\rho_{\mathcal{C}}$ is encoded in the path and polarization degrees of freedoms (DOFs) of photon~\cite{HUANG2022593,PhysRevLett.129.030502,Guo:18,PhysRevA.79.042101}. The state $\rho_\mathcal{S}$ encoded in the path is regulated by HWP1 and BD1. In each path, HWP2 and HWP3 can be configured to produce state $\rho_\mathcal{C}$ encoded in polarization. To facilitate the construction of the device for the evolution of quantum states, we convert path-polarization DOFs into only path DOF coding (subspace $|a\rangle$ of path a, and the like) through BD2.

Part $\mathbb{II}$: The construction of operator $\mathcal{E}$. When the optical axis of BD is laid in the horizontal plane, it refracts the horizontally polarized photon by 4~mm in the horizontal direction, and when the optical axis of BD is laid in the vertical plane, it refracts the vertically polarized photon by 4~mm in the vertical direction. BD3 coherently superimposes the path subspaces $|b\rangle$ and $|c\rangle$ on path c. HWP4 and HWP5 change the polarization of subspace $|b\rangle \oplus|c\rangle$ of path c and $|d\rangle$ of path d, respectively, and then the state is divided into two parts by PBS2. (1) The component of the horizontal polarization is transmitted by PBS2 and continues the evolution of $\mathcal{K}_1$. The function of BD5-BD7 is to implement a path switch to make the components $|c\rangle \to |a\rangle$ and $|d\rangle \to |c\rangle$. Then, the polarization of subspace $|c\rangle$ is changed by HWP7 and separated by BD8. Thus, the component of subspace $|c\rangle$ evolves to $|b\rangle $ and $|c\rangle$ after HWP7 and BD8. Then, the final state $\rho_{\mathcal{SC}}^{\mathcal{K}_1}$ of Kraus operator $\mathcal{K}_1$ enters PBS3. (2) The component of the vertical polarization is reflected by PBS2 and continues the evolution of $\mathcal{K}_0$, and the final state is $\rho_{\mathcal{SC}}^{\mathcal{K}_0}$. HWP6 and BD4 have the same functions as HWP7 and BD8. Through PBS3, we combine the evolution results ($\rho_{\mathcal{SC}}^{\mathcal{K}_0}$, $\rho_{\mathcal{SC}}^{\mathcal{K}_1}$). It is worth noting that after PBS3, $\rho_{\mathcal{SC}}^{\mathcal{K}_0}$ and $\rho_{\mathcal{SC}}^{\mathcal{K}_1}$ have different polarizations on each path. To fit the measurement setup, we select different HWPs-group for $\rho_{\mathcal{SC}}^{\mathcal{K}_0}$ and $\rho_{\mathcal{SC}}^{\mathcal{K}_1}$ to adjust the polarization of each path. If we replace PBS3 with a beamsplitter, then we can measure $\rho_{\mathcal{SC}}^{\mathcal{K}_0}$ and $\rho_{\mathcal{SC}}^{\mathcal{K}_1}$ simultaneously, which results in the loss of photon counts. More details of the evolution process can be found in section $\mathbf{II}$ of supplementary materials~\cite{SM}.

Part $\mathbb{III}$: Measurement setup. We use BD9 to encode the state in the path and polarization DOFs. The measurement device consists of HWP8-HWP10, QWPs, and BD10. This setup can realize tomography of the output state encoded on polarization and path DOFs.

Our goal is to implement the example given in the protocol, where $\rho_\mathcal{S}$ is a pure state and $\rho_\mathcal{C}$ is a mixed state. The mixed state can be decomposed into a probabilistic mixture of pure states~\cite{PhysRevLett.127.220501}. The quantum state $\rho_\mathcal{S} \otimes \rho_\mathcal{C}$ can be written as
\begin{equation}
    \rho_\mathcal{S} \otimes \rho_\mathcal{C}=0.7306 \cdot \rho_{\mathcal{S}}\otimes \rho_{c_0}+0.2694 \cdot \rho_{\mathcal{S}}\otimes \rho_{c_1},  \label{fenjie}
\end{equation}

We prepare these two pure states $\rho_{\mathcal{S}}\otimes \rho_{c_0}$ and $\rho_{\mathcal{S}}\otimes \rho_{c_1}$ in proportion and mix them. In the experiment, the total time of each measurement is 100~s, namely, the measurement duration of $\rho_{\mathcal{S}}\otimes \rho_{c_0}$ is 73.06~s, and the measurement duration of $\rho_{\mathcal{S}}\otimes \rho_{c_1}$ is 26.94~s.

{\itshape Results.---} In the experiment, we obtain the evolution process matrix $\mathbf{M}_p$ by process tomography, and the catalytic system is selected as the state (\ref{fuzhustate}). The tomography results of the system state and catalytic state in the experiment are:
\begin{align}
    \rho_{\mathcal{S}}^{exp}=\frac{1}{2}\left(I+0.4340\sigma_x+0.0219\sigma_y-0.8998\sigma_z\right), \\
    \rho_{\mathcal{C}}^{exp}=\frac{1}{2}\left(I+0.4363\sigma_x-0.0072\sigma_y+0.2912\sigma_z\right).
\end{align}

We obtain the evolved state of $\mathcal{S}$ and $\mathcal{C}$
\begin{align}
    \rho_{\mathcal{S}}^{\prime{\ }exp}=\frac{1}{2}\left(I+0.4597\sigma_x-0.0062\sigma_y-0.3301\sigma_z\right), \\
    \rho_{\mathcal{C}}^{\prime {\  }exp}=\frac{1}{2}\left(I+0.4800\sigma_x-0.0140\sigma_y+0.2776\sigma_z\right).
\end{align}

In each measurement, approximately 20,000 total events are generated per second. According to the protocol, we assure that the final catalytic state $\rho_{\mathcal{C}}^{\prime{\ }exp}$ can be transformed to the initial catalytic state $\rho_{\mathcal{C}}^{exp}$ through the TIO operation, which guarantees that asymmetric resources remain non-decreasing in system $\mathcal{C}$, while the asymmetry of system $\mathcal{S}$ increases significantly~\cite{Marvian_2013}. The asymmetric corrections for systems $\mathcal{S}$ and $\mathcal{C}$ are $\epsilon_{\mathcal{S}}=0.0080$ and $\epsilon_{\mathcal{C}}=0.0073$~\cite{SM}. Thus, Eq.(\ref{constraint}) is satisfied, and we can obtain the deterministic increase of the asymmetric resources of system $\mathcal{S}$ with $\Delta \eta = \robu(\rho_{\mathcal{S}}^{\prime {\ }exp})-\robu(\rho_\mathcal{S}^{exp})-\epsilon_{\mathcal{S}}=0.0172\pm0.0022$. This result confirms the achievement of an asymmetric resource amplification process based on catalytic-assisted states in the experiment. The standard deviation is estimated from Poissonian statistics of photon counting through 500 times Monte-Carlo simulations~\cite{harrison2010introduction,joy1991introduction}.

{\itshape Discussion.---} For the first time, on the heralded single-photon platform, we experimentally amplify the asymmetry of $\rho_{\mathcal{S}}$ under global TIO with the assistance of quantum catalyst $\rho_{\mathcal{C}}$ by reasonably correcting the impact of experimental errors. Although the experimental platform operates in a quantum regime, its behavior is equivalent to a classical one. Theoretically, our protocol is defined within the quantum framework as a quantum protocol that leverages quantum catalysts to amplify the quantum resource of quantum asymmetry, involving the evolution of quantum states through quantum operators. Asymmetric resources and their operations have not been defined within classical frameworks. Experimentally, we encoded systems $\mathcal{S}$ and $\mathcal{C}$ on the path and polarization DOFs of a single-photon. The quantum state preparation and quantum operator evolution throughout the entire quantum catalysis process utilized only the first-order coherence of photons~\cite{PhysRevA.82.033833,PhysRevLett.99.160401}, which is a property also observed in the classic regime. Practically, the resulting statistical distributions are identical to the intensity distributions obtained by using classical (coherent) light. 
If multiple photons or atoms are used as experimental platforms to encode systems $\mathcal{S}$ and $\mathcal{C}$, the quantum operations of the two systems require high-order coherence in photons, such as Hong-Ou-Mandel effect~\cite{OU1987118,doi:10.1126/sciadv.1500087}, controlled-NOT operation~\cite{doi:10.1126/science.abe3150}. In this case, the experimental results deviate from the classical behavior. Additionally, our protocol can be extended to multi-particle, multi-DOF physical systems. It is possible to use classical resources to replace quantum ones to simulate first-order coherence between multiple DOFs and to construct hybrid strategies for accelerating quantum tasks. We further demonstrate that the correlated catalytic protocol can be applied to amplify the asymmetry of additional quantum states. Simultaneously, we conducted simulations to assess the impact of a specific type of noise on the catalytic protocol in supplementary materials~\cite{SM}.

The catalyst state facilitates the manipulation and conversion of quantum resources. 
It has recently been shown that the correlated catalyst enables arbitrary manipulation of quantum coherence. However, because of the correlated catalyst, it shows a new type of resource appropriation, which is also discussed~\cite{PhysRevLett.128.240501,PhysRevLett.132.200201,PhysRevLett.132.180202}. The framework of dynamical resource theories, such as coherence, can also be used as a resource to determine the performance of Shor algorithms. It can bound the success probability of the algorithm~\cite{PhysRevLett.129.120501}. Therefore, we believe that the increase of asymmetry under coherent catalysis will also have significance for related fields, such as quantum computing, quantum coherent manipulation, and so on.

An intriguing future direction for quantum correlated catalysts is the relationship between the correlation introduced between the catalyst and the system. It changes in the system's resources, for which there is no clear conclusion currently. Exploring the potential in resource transformation and the efficient application of quantum catalysis can lay the foundation for the optimal utilization of quantum resources.

\begin{acknowledgments}
This work was supported by NSFC (No.~12374338, No.~11904357, No.~12174367,  No.~12204458, No.~11774205 and No.~62322513), the Innovation Program for Quantum Science and Technology (No. 2021ZD0301200), the Fundamental Research Funds for the Central Universities, China Postdoctoral Science Foundation (2021M700138), China Postdoctoral for Innovative Talents (BX2021289), the Shanghai Municipal Science and Technology Fundamental Project (No.~21JC1405400). This work was partially supported by the USTC Center for Micro and Nanoscale Research and Fabrication.
\end{acknowledgments}

\bibliography{apssamp_revise}

\end{document}


\title{SUPPLEMENTARY INFORMATION\\Experimental Catalytic Amplification of Asymmetry}
\author{Chao Zhang}
\email{These two authors contributed equally to this work.}
\affiliation{CAS Key Laboratory of Quantum Information, University of Science and Technology of China, Hefei, 230026, China}
\affiliation{CAS Center For Excellence in Quantum Information and Quantum Physics, University of Science and Technology of China, Hefei 230026, China}

\author{Xiao-Min Hu}
\email{These two authors contributed equally to this work.}
\affiliation{CAS Key Laboratory of Quantum Information, University of Science and Technology of China, Hefei, 230026, China}
\affiliation{CAS Center For Excellence in Quantum Information and Quantum Physics, University of Science and Technology of China, Hefei 230026, China}

\author{Feng Ding}
\affiliation{School of Information Science and Engineering, Shandong University, Qingdao 266237, China}

\author{Xue-Yuan Hu}
\email{xyhu@sdu.edu.cn}
\affiliation{School of Information Science and Engineering, Shandong University, Qingdao 266237, China}

\author{Yu Guo}
\affiliation{CAS Key Laboratory of Quantum Information, University of Science and Technology of China, Hefei, 230026, China}
\affiliation{CAS Center For Excellence in Quantum Information and Quantum Physics, University of Science and Technology of China, Hefei 230026, China}

\author{Bi-Heng Liu}
\email{bhliu@ustc.edu.cn}
\affiliation{CAS Key Laboratory of Quantum Information, University of Science and Technology of China, Hefei, 230026, China}
\affiliation{CAS Center For Excellence in Quantum Information and Quantum Physics, University of Science and Technology of China, Hefei 230026, China}
\affiliation{Hefei National Laboratory, University of Science and Technology of China, Hefei 230088, China}

\author{Yun-Feng Huang}
\affiliation{CAS Key Laboratory of Quantum Information, University of Science and Technology of China, Hefei, 230026, China}
\affiliation{CAS Center For Excellence in Quantum Information and Quantum Physics, University of Science and Technology of China, Hefei 230026, China}
\affiliation{Hefei National Laboratory, University of Science and Technology of China, Hefei 230088, China}

\author{Chuan-Feng Li}
\affiliation{CAS Key Laboratory of Quantum Information, University of Science and Technology of China, Hefei, 230026, China}
\affiliation{CAS Center For Excellence in Quantum Information and Quantum Physics, University of Science and Technology of China, Hefei 230026, China}
\affiliation{Hefei National Laboratory, University of Science and Technology of China, Hefei 230088, China}

\author{Guang-Can Guo}
\affiliation{CAS Key Laboratory of Quantum Information, University of Science and Technology of China, Hefei, 230026, China}
\affiliation{CAS Center For Excellence in Quantum Information and Quantum Physics, University of Science and Technology of China, Hefei 230026, China}
\affiliation{Hefei National Laboratory, University of Science and Technology of China, Hefei 230088, China}

\maketitle
\clearpage
\onecolumngrid

\section{Discussion on the effects of unavoidable noises}

Catalysis extends the methods available for manipulating quantum resources. Whether the implemented catalysis protocol is quantum can be determined based on the definition in \cite{datta2023catalysis}: `Quantum catalysis is conceptually similar to chemical catalysis but differs from it in several important details. A simple analogy between quantum and chemical catalysis can be established by replacing “chemical reaction” with “quantum state transition”. With this, a quantum catalyst is a quantum system which enables otherwise impossible transitions between quantum states.'.

In our protocol, with the auxiliary quantum state 
$\rho_c=\frac{1}{2}\left(\mathbb{I}+0.5710\sigma_{x}+0.2928\sigma_{z}\right)$ remaining unchanging, the initial quantum state
\begin{align}
    \rho_s=\frac{1}{2}(\iden+0.4333\sigma_{x}-0.9013\sigma_{z}) \notag
\end{align}
of system $\mathcal{S}$ is transformed into 
\begin{equation}
\rho^{\prime}_s=\frac{1}{2}\left(\iden+0.5314 \sigma_{x}-0.3251 \sigma_{z}\right)\notag
\end{equation}
through global TIO $\mathcal{E}$. This transformation amplifies the quantum asymmetric resources of system $\mathcal{S}$, which would be impossible without the catalyst $\rho_c$. Thus, from the theoretical perspective, we construct a quantum catalysis protocol to amplify quantum asymmetric resources, which is likely to enhance the performance of related applications. However, as with all quantum catalytic protocols, its execution is subject to experimental noise. There are two imperfections we have to take into account due to the unavoidable noise in our experiment. 


1. The realizable completely positive and trance-preserving (CPTP) map on the composed system may deviate from translationally invariant operation (TIO). Thus, subtracting the additional asymmetric resources introduced by experimental noise into systems $\mathcal{S}$ and $\mathcal{C}$.

2. The output catalyst state may not be exactly the same as its initial state. Thus, comparing the initial state with the final state of system $\mathcal{C}$ ensures that the asymmetric resources of the latter remain non-decreasing. This implies that free operations can revert the final state of system $\mathcal{C}$ to its initial state, preventing the misappropriation of asymmetric resources within system $\mathcal{C}$.

These constraints ensure that the increment of the asymmetry that we observe in the experiment is indeed caused by the catalytic transformation, instead of noise effects. Firstly, to estimate the error caused by the deviation of the realizable CPTP map $\cptp$ from TIO, we use the Choi-Jamiolkowski (CJ) presentation of quantum channels. The CJ matrix of a quantum channel $\cptp$ reads
\begin{equation}
J_\cptp=\left[\begin{array}{cccc}
\cptp(|0\rangle\langle0|) & \cptp(|0\rangle\langle1|) & \cdots & \cptp(|0\rangle\langle d|)\\
\cptp(|1\rangle\langle0|) & \cptp(|1\rangle\langle1|) & \cdots & \cptp(|1\rangle\langle d|)\\
\vdots & \vdots & \ddots & \vdots\\
\cptp(|d\rangle\langle0|) & \cptp(|d\rangle\langle1|) & \cdots & \cptp(|d\rangle\langle d|)
\end{array}
\right],
\end{equation}
where $|i\rangle$ are eigenstates of the system on which $\cptp$ acts. In our case, $\cptp$ acts on the composed system of $\mathcal{S}\otimes \mathcal{C}$, and hence, $d=3$ and $|0\rangle=|0_\mathcal{S}0_\mathcal{C}\rangle$, $|1\rangle=|0_\mathcal{S}1_\mathcal{C}\rangle$, $|2\rangle=|1_\mathcal{S}0_\mathcal{C}\rangle$, $|3\rangle=|1_\mathcal{S}1_\mathcal{C}\rangle$. Now, we introduce the following matrix
\begin{equation}
M=\left[\begin{array}{cccc}
M_0 & M_1 & M_1 & M_2\\
M_1^\dagger & M_0 & M_0 & M_1\\
M_1^\dagger & M_0 & M_0 & M_1\\
M_2^\dagger & M_1^\dagger & M_1^\dagger & M_0
\end{array}
\right]
\end{equation}
with
\begin{eqnarray}
M_0=\left[\begin{array}{cccc}
1 & 0 & 0 & 0\\
0 & 1 & 1 & 0\\
0 & 1 & 1 & 0\\
0 & 0 & 0 & 1
\end{array}
\right],
M_1=\left[\begin{array}{cccc}
0 & 1 & 1 & 0\\
0 & 0 & 0 & 1\\
0 & 0 & 0 & 1\\
0 & 0 & 0 & 0
\end{array}
\right],\nonumber
M_2=\left[\begin{array}{cccc}
0 & 0 & 0 & 1\\
0 & 0 & 0 & 0\\
0 & 0 & 0 & 0\\
0 & 0 & 0 & 0
\end{array}
\right].
\end{eqnarray}
As proven in Ref. \cite{PhysRevA.103.022403}, a CPTP map belongs to TIO iff
\begin{equation}
J_\cptp=J_\cptp\odot M,
\end{equation}
where $\odot$ denotes the Hadamard product, i.e., the entrywise matrix product. Furthermore, if $\cptp_{exp}$ in experiment does not belong to TIO, it is directly checked that $J_{\cptp_\mathrm{TI}}\equiv J_\cptp\odot M$ is the CJ matrix of a translationally invariant operation $\cptp_\mathrm{TI}$. The errors, which are caused by the deviation of $\cptp_{exp}$ from $\cptp_\mathrm{TI}$, on systems $\mathcal{S}$ and $\mathcal{C}$, can then be defined as
\begin{equation}
\epsilon_\mathcal{S}=||\Phi_S||_1, \epsilon_\mathcal{C}=||\Phi_C||_1,  \label{wucha}
\end{equation}
where $\Phi_S(\cdot)=\tr_\mathcal{C}\circ(\cptp_{exp}-\cptp_\mathrm{TI})[\cdot\otimes\rho_C]$ and $\Phi_C(\cdot)=\tr_\mathcal{S}\circ(\cptp_{exp}-\cptp_\mathrm{TI})[\rho_S\otimes\cdot]$. Here $\circ$ denotes the composition of quantum channels, $\rho_S$ and $\rho_C$ are respectively the input states of $\mathcal{S}$ and $\mathcal{C}$, and $||\cdot||_1$ is the diamond norm. This means that, the distance between the output states of $\mathcal{C}$ from channel $\tr_\mathcal{S}\circ\cptp_{exp}[\rho_S\otimes\cdot]$ and from its TI counterpart $\tr_\mathcal{S}\circ\cptp_\mathrm{TI}$, is upper bounded by $\epsilon_\mathcal{C}$. 

Secondly, let the input states of $\mathcal{S}$ and $\mathcal{C}$ be $\rho_\mathcal{S}=\frac12(\iden+\eta\sigma_x+\mu\sigma_y+\xi\sigma_z)$ and $\rho_C=\frac12(\iden+x\sigma_x+y\sigma_y+z\sigma_z)$, and the corresponding output states be $\rho_\mathcal{S}'=\frac12(\iden+\eta'\sigma_x+\delta\sigma_y+\xi'\sigma_z$) and $\rho'_C=\frac12(\iden+x'\sigma_x+y'\sigma_y+z'\sigma_z)$. Then, the first condition, namely, $\rho_\mathcal{C}\in\tiocone(\rho'_\mathcal{C})$, is equivalent to $\sqrt{x^2+y^2}\max\{\sqrt{\frac{1-z'}{1-z}},\sqrt{\frac{1+z'}{1+z}}\}\leq\sqrt{x^{\prime2}+y^{\prime2}}$. Because $|x'|\leq\sqrt{x^{\prime2}+y^{\prime2}}$, it is sufficient to require $\mathrm{lhs}\leq|x'|$. Considering the error $\epsilon_\mathcal{C}$, the sufficient condition for $\rho_\mathcal{C}\in\tiocone(\rho_\mathcal{C}^\mathrm{TI})$ reads
\begin{equation}
\sqrt{x^2+y^2}\max\left\{\sqrt{\frac{1-z'}{1-z}},\sqrt{\frac{1+z'}{1+z}}\right\}\leq |x'|-\epsilon_\mathcal{C}.  \label{yueshu}
\end{equation}
By applying similar discussions to $\mathcal{S}$, we modify the increment in the robustness of asymmetry in $\mathcal{S}$ to $\Delta\tilde{\eta}=\robu(\rho^{\prime}_{\mathcal{S}})-\robu(\rho_\mathcal{S})-\epsilon_\mathcal{S}$.

\section{Details of constructing channel $\mathcal{E}_{exp}$}
Usually, the implementation of quantum channels can be achieved by introducing environmental system $\mathbf{E}$ to perform global unitary evolution, and tracing system $\mathbf{E}$ to obtain the final result. Also, the quantum channel is known as the operator-sum representation~\cite{doi:10.1126/science.1139892}, i.e. Kraus operation. The channel $\mathcal{E}$ in the experiment consists of two Kraus operators $\mathcal{K}_0$ and $\mathcal{K}_1$, where
\begin{equation}
\mathcal{K}_{0}=\left[\begin{array}{cccc}
1 & 0 & 0 & 0 \\
0 & -0.1573 & 0.4029 & 0 \\
0 & -0.3278 & 0.8400 & 0 \\
0 & 0 & 0 & 0.7445
\end{array}\right],    
\mathcal{K}_{1}=\left[\begin{array}{cccc}
0 & \ \ \,0.9315 & 0.3636 & 0 \\
0 & 0 & 0 & 0.4721 \\
0 & 0 & 0 & 0.4721 \\
0 & 0 & 0 & 0
\end{array}\right].   \label{Kraus}
\end{equation}
Note that there are many zero values in the operator, and it has the characteristic of $\frac{\mathcal{K}_0^{3,2}}{\mathcal{K}_0^{2,2}}=\frac{\mathcal{K}_0^{3,3}}{\mathcal{K}_0^{2,3}}$, where $\mathcal{K}_0^{i,j}$ represents the elements in the i-th row and j-th column of the matrix $\mathcal{K}_0$. In the experiment, we chose path degrees of freedom (DOFs) as the basis $\{|a\rangle, |b\rangle, |c\rangle, |d\rangle\}$, and with the assistance of polarization DOF on each path, we constructed channel $\mathcal{E}$ specially. The detailed process is as follows.

\begin{figure*}[tph!]
\includegraphics [width=0.9\textwidth]{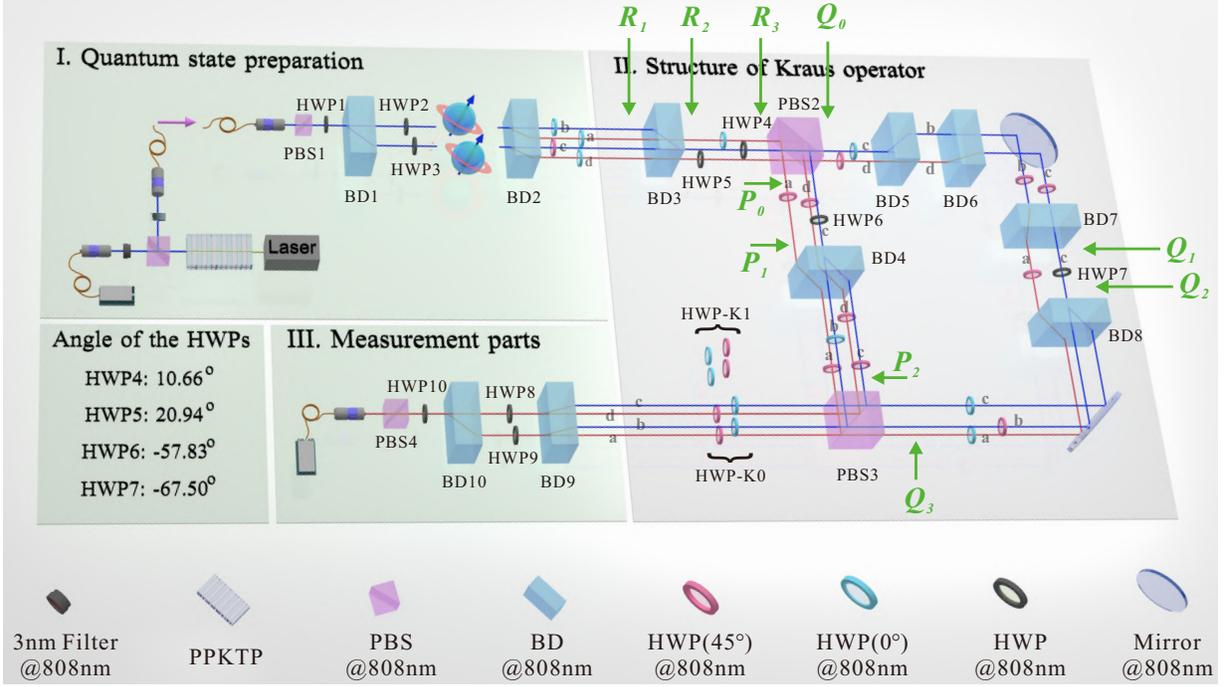}
\vspace{-0.3cm}%
\caption{$\mathbb{I}$: Quantum state preparation. $\mathbb{II}$: Structure of Kraus $\mathcal{K}_0$ and $\mathcal{K}_1$. The letters (a, b, c, d) on the ray represent the encoding of the path, where the upper layer of ray (b, c) is colored blue, and the lower layer of ray (a, d) is colored red. The quantum states reflected and transmitted by PBS3, denoted as $\rho_{\mathcal{SC}}^{\mathcal{K}_0}$ and $\rho_{\mathcal{SC}}^{\mathcal{K}_1}$, respectively, correspond to the final states resulting from the evolution of Kraus operators $\mathcal{K}_0$ and $\mathcal{K}_1$. Notably, these states exhibit different polarizations. To be consistent with the polarization on each path before BD9 with $|a_H\rangle, |b_V\rangle, |c_V\rangle, |d_H\rangle$, where the subscript $H(V)$ stands for horizontal (vertical) polarization in the path, the angles of HWPs-group after PBS3 need to be adjusted. As shown in the figure, the HWPs-group named HWP-K0 (HWP-K1) adjusts the polarization of state $\rho_{\mathcal{SC}}^{\mathcal{K}_0}$  ($\rho_{\mathcal{SC}}^{\mathcal{K}_1}$). The angles of HWP4-HWP7 used to construct the channel are also given in detail. $\mathbb{III}$: Measurement parts.}
\label{fig:setup}
\vspace{0.2cm}  \label{gouzao}
\end{figure*}

In the description process, we represent the polarization along the corresponding path coding as subscripts in quantum states. Without loss of generality, we assume that the quantum state prepared is recorded in path DOF ('a', 'b', 'c', 'd') by using BD2, and the state can be rewritten as
\begin{equation}
|\phi\rangle=x_a|a_V\rangle+x_b|b_H\rangle+x_c|c_H\rangle+x_d|d_V\rangle
\end{equation}
after BD2, where $\{x_a, x_b, x_c, x_d\}$ represent the amplitudes of subspace $\{|a\rangle, |b\rangle, |c\rangle, |d\rangle\}$ and satisfy $|x_a|^2+|x_b|^2+|x_c|^2+|x_d|^2=1$. The subscript $'H(V)'$ indicates that the photon is horizontally (vertically) polarized in this path, i.e., we utilized the polarization degree of freedom in each path. Set the angles of HWP4$\sim$HWP7 to $\theta_4\sim\theta_7$, respectively. Next, we present the quantum states during the evolution process at certain positions in the structure, consistent with the symbols in Figure \ref{gouzao}.$\mathbb{II}$.

$R_1$: 
\begin{equation}
|\phi_{R_1}\rangle=-x_a|a_V\rangle+x_b|b_H\rangle+x_c|c_V\rangle-x_d|d_V\rangle , \label{R1}
\end{equation}

$R_2$:
\begin{equation} |\phi_{R_2}\rangle=-x_a|a_V\rangle+x_b|c_H\rangle+x_c|c_V\rangle-x_d|d_V\rangle , \label{R2}
\end{equation}

$R_3$:
\begin{equation}
|\phi_{R_3}\rangle=x_a|a_V\rangle+(x_b\cos{2\theta_4}+x_c\sin{2\theta_4})|c_H\rangle+(x_b\sin{2\theta_4}-x_c\cos{2\theta_4})|c_V\rangle-x_d\sin{2\theta_5}|d_H\rangle+x_d\cos{2\theta_5}|d_V\rangle , \label{R3}
\end{equation}

$P_0$:
\begin{equation}
|\phi_{P_0}\rangle=x_a|a_V\rangle+(x_b\sin{2\theta_4}-x_c\cos{2\theta_4})|c_V\rangle+x_d\cos{2\theta_5}|d_V\rangle , \label{P0}
\end{equation}

$P_1$:
\begin{equation}
|\phi_{P_1}\rangle=x_a|a_H\rangle+(x_b\sin{2\theta_4}-x_c\cos{2\theta_4})\sin{2\theta_6}|c_H\rangle-(x_b\sin{2\theta_4}-x_c\cos{2\theta_4})\cos{2\theta_6}|c_V\rangle+x_d\cos{2\theta_5}|d_H\rangle , \label{P1}
\end{equation}

$P_2$:
\begin{equation}
|\phi_{P_2}\rangle=x_a|a_V\rangle+(x_b\sin{2\theta_4}-x_c\cos{2\theta_4})\cos{2\theta_6}|b_V\rangle+(x_b\sin{2\theta_4}-x_c\cos{2\theta_4})\sin{2\theta_6}|c_V\rangle+x_d\cos{2\theta_5}|d_V\rangle, \label{P2}
\end{equation}

$Q_0$:
\begin{equation}
|\phi_{Q_0}\rangle=(x_b\cos{2\theta_4}+x_c\sin{2\theta_4})|c_H\rangle-x_d\sin{2\theta_5}|d_H\rangle,  \label{Q0}
\end{equation}

$Q_1$:
\begin{equation}
|\phi_{Q_1}\rangle=(x_b\cos{2\theta_4}+x_c\sin{2\theta_4})|a_V\rangle-x_d\sin{2\theta_5}|c_H\rangle,  \label{Q1}
\end{equation}

$Q_2$:
\begin{equation}
|\phi_{Q_2}\rangle=(x_b\cos{2\theta_4}+x_c\sin{2\theta_4})|a_H\rangle-x_d\sin{2\theta_5}\cos{2\theta_7}|c_H\rangle-x_d\sin{2\theta_5}\sin{2\theta_7}|c_V\rangle , \label{Q2}
\end{equation}

$Q_3$:
\begin{equation}
|\phi_{Q_3}\rangle=(x_b\cos{2\theta_4}+x_c\sin{2\theta_4})|a_H\rangle-x_d\sin{2\theta_5}\cos{2\theta_7}|b_H\rangle-x_d\sin{2\theta_5}\sin{2\theta_7}|c_H\rangle.  \label{Q3}
\end{equation}
Based on the above processes, for any initial state $|\phi\rangle$ with path DOF coding, i.e. $(x_a;x_b;x_c;x_d)$, follow the evolutionary order of $R_1, R_2, R_3, P_0, P_1, P_2$, and the final state is $|\phi_{P_2}\rangle$, i.e., with evolution $G_{RP}$:
\begin{equation}
    G_{RP}\left(\begin{array}{c}
x_a  \\
x_b  \\
x_c  \\
x_d 
\end{array}\right)=\left(\begin{array}{c}
x_a  \\
(x_b\sin{2\theta_4}-x_c\cos{2\theta_4})\cos{2\theta_6}  \\
(x_b\sin{2\theta_4}-x_c\cos{2\theta_4})\sin{2\theta_6}  \\
x_d\cos{2\theta_5}
\end{array}\right) . \label{K0}
\end{equation}
Formula (\ref{K0}) holds for any initial state $|\phi\rangle$, so it is easy to obtain
\begin{equation}
G_{RP}=\left[\begin{array}{cccc}
1 & 0 & 0 & 0 \\
0 & \sin{2\theta_4}\cos{2\theta_6} & -\cos{2\theta_4}\cos{2\theta_6} & 0 \\
0 & \sin{2\theta_4}\sin{2\theta_6} & -\cos{2\theta_4}\sin{2\theta_6} & 0 \\
0 & 0 & 0 & \cos{2\theta_5}
\end{array}\right] . \label{GRP}
\end{equation}
Similarly, for any initial state $|\phi\rangle$, following the evolutionary order $R_1, R_2, R_3, Q_0, Q_1, Q_2$, the final state is $|\phi_{Q_3}\rangle$, and the corresponding evolution $G_{RQ}$ is:
\begin{equation}
G_{RQ}=\left[\begin{array}{cccc}
0 & \cos{2\theta_4} & \sin{2\theta_4} & 0 \\
0 & 0 & 0 & -\sin{2\theta_5}\cos{2\theta_7} \\
0 & 0 & 0 & -\sin{2\theta_5}\sin{2\theta_7} \\
0 & 0 & 0 & 0
\end{array}\right].  \label{GRQ}
\end{equation}
Based on the above processes, we provide the angle of HWP4$\sim$HWP7, as shown in Figure \ref{gouzao}. Substituting them into Eq. (\ref{GRP}) and (\ref{GRQ}) can construct the target Kraus operator (\ref{Kraus}). Even though we have specially constructed the Kraus operator (\ref{Kraus}), there are still imperfections in the experimental implementation. Therefore, we need to provide a detailed characterization of the implemented channel $\mathcal{E}_{exp}$ and eliminate its potential impact.

\section{Tomography of state and Process Tomography Of Channel}\label{aaaaaaa}

Taking a tomography approach~\cite{PhysRevA.66.012303} to the output state, we obtain the density matrix of the system $\mathcal{S}\otimes \mathcal{C}$ and their subspace. For system $\mathcal{S}$ encoded in path DOF, there is a complete set of measurement bases:

\begin{align}
    &|\phi_{\mathcal{S}1}\rangle=\left( \begin{array}{c}
     1     \\
     0  
    \end{array}\right), \notag
    &|\phi_{\mathcal{S}2}\rangle=\left( \begin{array}{c}
     0     \\
     1  
    \end{array}\right),  \\   
    &|\phi_{\mathcal{S}3}\rangle=\frac{1}{\sqrt{2}}\left( \begin{array}{c}
     1     \\
     1  
    \end{array}\right),  
    &|\phi_{\mathcal{S}4}\rangle=\frac{1}{\sqrt{2}}\left( \begin{array}{c}
     1     \\
     i  
    \end{array}\right),   \label{Sji}
\end{align}
with $|\phi_{S1}\rangle\langle\phi_{S1}|+|\phi_{S2}\rangle\langle\phi_{S2}|=\mathbf{I}$. Similarly, for system $\mathcal{C}$ encoded in the polarization DOF, there also is a complete set of measurement bases:
\begin{align}
    &|\phi_{\mathcal{C}1}\rangle=\left( \begin{array}{c}
     1     \\
     0  
    \end{array}\right), \notag
    &|\phi_{\mathcal{C}2}\rangle=\left( \begin{array}{c}
     0     \\
     1  
    \end{array}\right),  \\  
    &|\phi_{\mathcal{C}3}\rangle=\frac{1}{\sqrt{2}}\left( \begin{array}{c}
     1     \\
     1  
    \end{array}\right),
    &|\phi_{\mathcal{C}4}\rangle=\frac{1}{\sqrt{2}}\left( \begin{array}{c}
     1     \\
     i  
    \end{array}\right), \label{Cji}
\end{align}
with $|\phi_{\mathcal{C}1}\rangle\langle\phi_{\mathcal{C}1}|+|\phi_{\mathcal{C}2}\rangle\langle\phi_{\mathcal{C}2}|=\mathbf{I}$. We perform the measurement based on the path-polarization DOF with $|\phi_k\rangle\langle\phi_k|=|\phi_{\mathcal{S}i}\rangle\langle\phi_{\mathcal{S}i}|\otimes|\phi_{\mathcal{C}j}\rangle\langle\phi_{\mathcal{C}j}|$ on the system $\mathcal{S}\otimes \mathcal{C}$, where $k=4(i-1)+j$, $i,j\in \left\{1,2,3,4\right\}$. Thus, when we perform the tomography on the system $\mathcal{S}$ in the whole space, we choose the complete measurement based on path DOF (\ref{Sji}) in system $\mathcal{S}$, i.e., in path-polarization DOFs in system $\mathcal{S}\otimes \mathcal{C}$ with $|\phi_{\mathcal{S}i}\rangle\langle\phi_{\mathcal{S}i}|\otimes\mathbf{I}=|\phi_{\mathcal{S}i}\rangle\langle\phi_{\mathcal{S}i}|\otimes(|\phi_{\mathcal{C}1}\rangle\langle\phi_{\mathcal{C}1}|+|\phi_{\mathcal{C}2}\rangle\langle\phi_{\mathcal{C}2}|)$; Thus, the choice of measurement bases are $|\phi_{\mathcal{S}i}\rangle\otimes|\phi_{\mathcal{C}1}\rangle$ and $|\phi_{\mathcal{S}i}\rangle\otimes|\phi_{\mathcal{C}2}\rangle$, $i\in\left\{1,2,3,4\right\}$. Similarly, when we perform the tomography on the system $\mathcal{C}$ in the whole space, we choose the complete measurement based on polarization DOF (\ref{Cji}) in system $\mathcal{C}$, i.e., in the path-polarization DOF in system $\mathcal{S}\otimes \mathcal{C}$ with $\mathbf{I}\otimes|\phi_{\mathcal{C}i}\rangle\langle\phi_{\mathcal{C}i}|=(|\phi_{\mathcal{S}1}\rangle\langle\phi_{\mathcal{S}1}|+|\phi_{\mathcal{S}2}\rangle\langle\phi_{\mathcal{S}2}|)\otimes|\phi_{\mathcal{C}j}\rangle\langle\phi_{\mathcal{C}j}|$; Thus, the choice of measurement bases are $|\phi_{\mathcal{S}1}\rangle\otimes|\phi_{\mathcal{C}j}\rangle$ and $|\phi_{\mathcal{S}2}\rangle\otimes|\phi_{\mathcal{C}j}\rangle$, $j\in\{1,2,3,4\}$.

Imperfect TIO operations $\mathcal{E}_{exp}$ can also introduce additional asymmetry into the system. To accurately describe the channel, we perform process tomography~\cite{PhysRevA.68.012305} on the channel, and the specific steps are as follows. For a complete set of measurement bases
\begin{align}
    |&\psi_1\rangle=\left(1,0,0,0\right)^{\prime}, \quad\quad\quad\quad\notag |\psi_2\rangle=\left(0,1,0,0\right)^{\prime}, \\ \notag
    |&\psi_3\rangle=\left(0,0,1,0\right)^{\prime}, \quad\quad\quad\quad\,\notag   |\psi_4\rangle=\left(0,0,0,1\right)^{\prime}, \\\notag
    |&\psi_5\rangle=\left(1,1,0,0\right)^{\prime}/\sqrt{2}, \quad\quad\,\,\notag |\psi_6\rangle=\left(1,i,0,0\right)^{\prime}/\sqrt{2}, \\\notag
    |&\psi_7\rangle=\left(1,0,1,0\right)^{\prime}/\sqrt{2}, \quad\quad\,\,\notag |\psi_8\rangle=\left(1,0,i,0\right)^{\prime}/\sqrt{2}, \\\notag 
    |&\psi_9\rangle=\left(0,1,1,0\right)^{\prime}/\sqrt{2},\quad\quad\,\,\notag
    |\psi_{10}\rangle=\left(0,1,i,0\right)^{\prime}/\sqrt{2}, \\\notag 
    |&\psi_{11}\rangle=\left(1,0,0,1\right)^{\prime}/\sqrt{2}, \quad\quad\notag
    |\psi_{12}\rangle=\left(1,0,0,i\right)^{\prime}/\sqrt{2}, \\\notag 
    |&\psi_{13}\rangle=\left(0,1,0,1\right)^{\prime}/\sqrt{2}, \quad\quad\notag  |\psi_{14}\rangle=\left(0,1,0,i\right)^{\prime}/\sqrt{2}, \\\notag 
    |&\psi_{15}\rangle=\left(0,0,1,1\right)^{\prime}/\sqrt{2}, \quad\quad\notag  |\psi_{16}\rangle=\left(0,0,1,i\right)^{\prime}/\sqrt{2}. \\ \label{basis_1}
\end{align}
 The symbol $^\prime$ represents just the transpose without the conjugate of a matrix. Then the state corresponding to each basis vector in (\ref{basis_1}) is evolved through the channel $\mathcal{E}$, and the output state is analyzed by tomography in whole space. Thus the process matrix can be reconstructed according to the method in \cite{PhysRevA.68.012305}. According to the precise description of $\mathbf{M}_p$ which deviates from TIO, the influence on each system can be analyzed. The detailed values of the process matrix $\mathbf{M}_p$ are listed in Table~\ref{real} and Table~\ref{imag}. We achieve a high fidelity $F_{process matrix}=99.04\pm0.05\%$.
 
\section{Selection of auxiliary state $\rho_\mathcal{C}$}  

In the main manuscript, it is mentioned that the asymmetry increase of the main system $\mathcal{S}$ can obtain 0.0982, where the system $\mathcal{S}$ reads
\begin{align}
    \rho_\mathfrak{S}=\frac{1}{2}(\iden+0.4333\sigma_{x}-0.9013\sigma_{z}),
\end{align}
and the corresponding catalytic state $\rho_\mathfrak{C}$ reads
\begin{equation}
\rho_\mathfrak{C}=\frac{1}{2}\left(\mathbb{I}+0.5710 \sigma_{x}+0.2928\sigma_{z}\right) .\label{fuzhustate}
\end{equation}

Nevertheless, the imperfection of channel $\cptp_{exp}$ will result in the catalytic state (\ref{fuzhustate}) not satisfying the constraint (\ref{yueshu}) after evolution. Based on the channel $\mathcal{E}_{exp}$ with process matrix $\mathbf{M}_p$ obtained by process tomography, the catalytic state is modified as follows to study the asymmetry increment of system $\mathcal{S}$:
\begin{equation}
\rho_{\mathcal{C}}=\frac{1}{2}\left(\mathbb{I}+(0.5710+\delta x) \sigma_{x}+(0.2928+\delta z)\sigma_{z}\right) .
\label{fuzhustate_xiuzheng}
\end{equation}

\begin{figure*}[tph!]
\includegraphics [width= 0.6\textwidth]{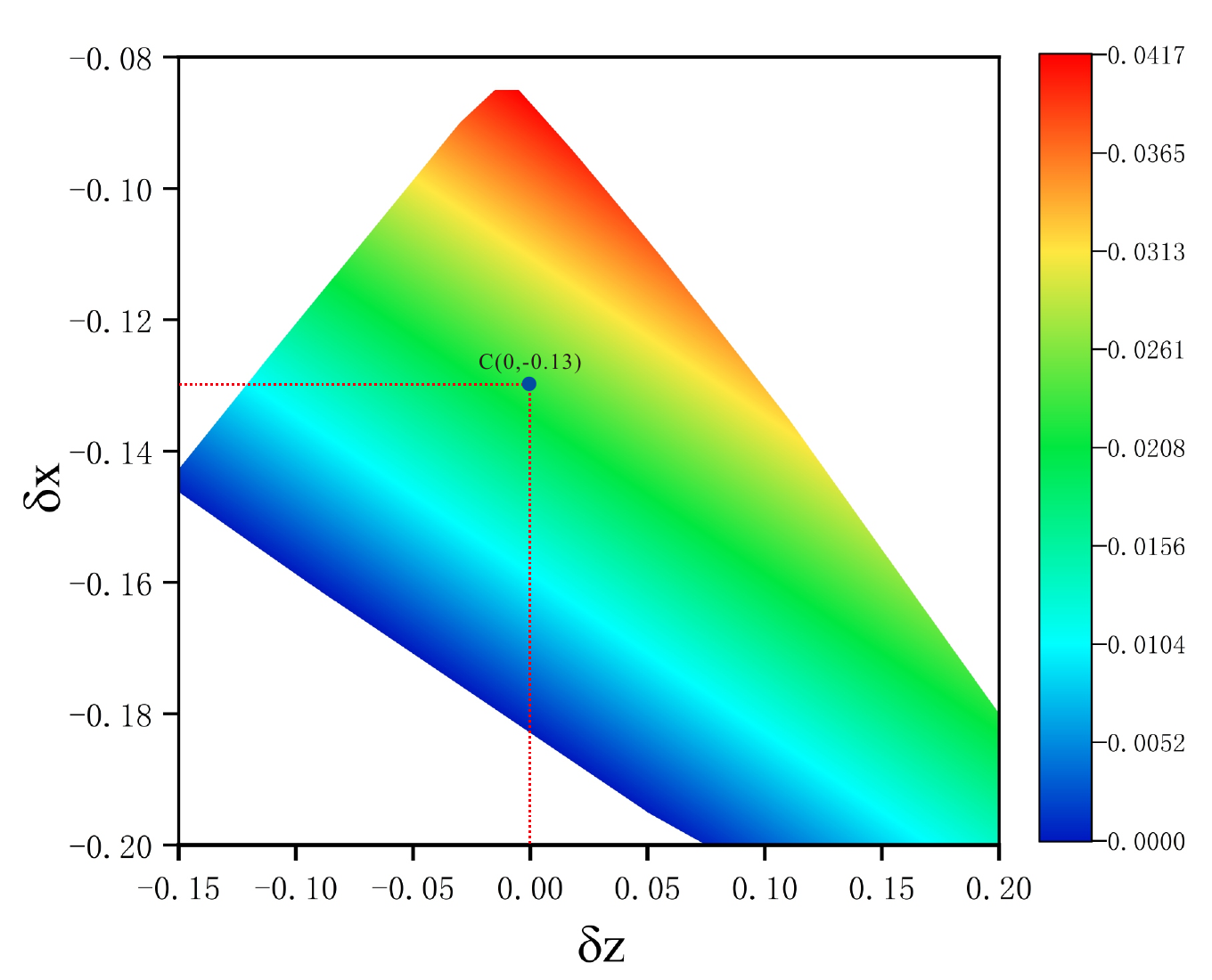}
\vspace{-0.3cm}%
\caption{The range of catalytic states is optional under the current channel $\mathbf{M}_p$ and the effective asymmetry increase of system $\mathcal{S}$. The auxiliary state in this region still satisfies constraint (\ref{yueshu}) after evolution with imperfect channel $\mathcal{E}_{exp}$ with process matrix $\mathbf{M}_p$, and system $\mathcal{S}$ has an effective asymmetry increase $\Delta\eta>0$, which is represented by the colors in the figure. The blue dot represents the target auxiliary state we will prepare. }
\vspace{0.2cm}  
\label{figureS1}
\end{figure*}

As shown in Figure \ref{figureS1}, the asymmetry increase of system $\mathcal{S}$ under the current channel is $\Delta\eta=0.0417$ at most, which is greatly reduced compared with the ideal value of 0.0982. Considering the deviation of state preparation, we conservatively select the correction value $(\sigma_z, \sigma_x)=(0, -0.13)$(blue dot) for $\rho_\mathcal{C}$, at which point the system $\mathcal{S}$ still has an objective asymmetry increase.

 \section{Consideration of Catalytic Amplification of Asymmetry of General Quantum State}

 In practical applications and experiments, quantum states are more diverse, so here we consider general qubit quantum states of system $\mathcal{S}$:
\begin{equation}
    \rho=\frac{1}{2}(I+r\sin{\theta}\cos{\phi}\sigma_x+r\sin{\theta}\sin{\phi}\sigma_y+r\cos{\theta}\sigma_z),
\end{equation}
where $r\in[0,1]$, $\theta\in[0,\pi]$, and $\phi\in[0,2\pi)$. By using free unitary $U=\diag\left[1, e^{-i\phi}\right]$, the above general qubit state can be transformed to the following form
\begin{equation}
    \rho_s=\frac{1}{2}(I+x\sigma_x+z\sigma_z), \label{state}
\end{equation}
where $x=r\sin{\theta}$ and $z=r\cos{\theta}$. Further, states in the form of $\rho_s$ can also be transformed back to $\rho$ via $U^{\dagger}$. Therefore, the maximal catalytic increment of asymmetry for state $\rho$ is the same as that for state $\rho_s$. Besides, in the calculation of the maximal catalytic increment of asymmetry, the result remains unchanged when one exchanges the states $|0\rangle$ and $|1\rangle$ in the computational basis. It follows that the increment for state $\rho'_s=\frac{1}{2}(I+x\sigma_x-z\sigma_z)$ is the same as that for $\rho_s$. Therefore, we only need to calculate the maximal increment for states in Eq. (\ref{state}) with $\theta\in[0,\pi/2]$.

\begin{figure*}[tph!]
\includegraphics [width=1\textwidth]{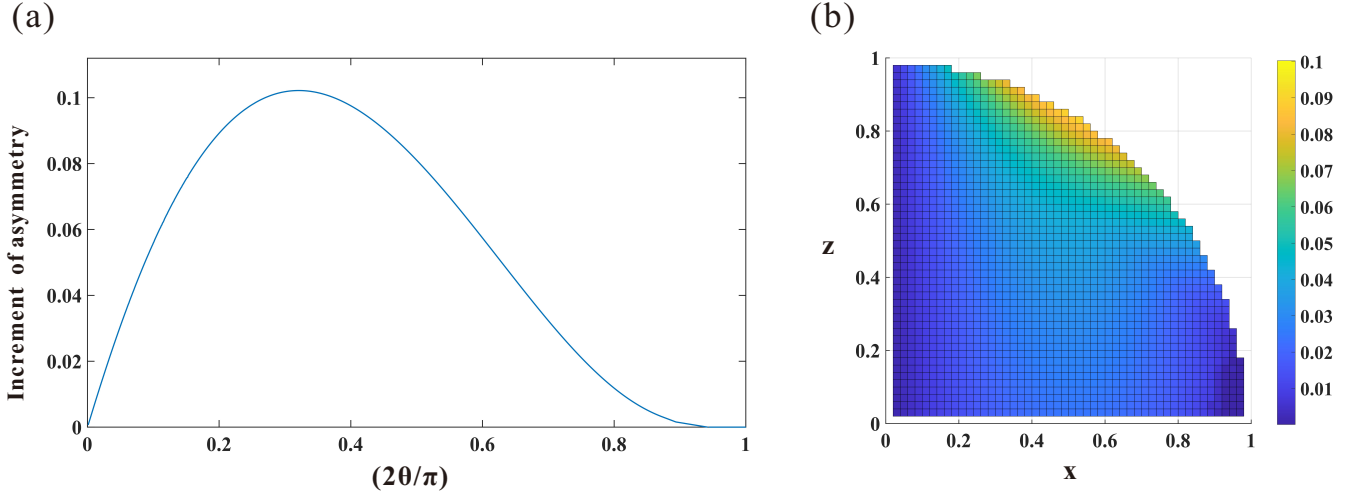}
\vspace{-0.3cm}%
\caption{ (a) Consider the pure state of formula (\ref{state}), i.e., $x^2+z^2=1$. In the horizontal axis, $\theta$ represents the angle between the block vector of the quantum state $\rho_s$ and the z-axis. The vertical axis represents the increment of asymmetry. For all pure states with $\theta\neq0,\pi/2$, theoretically, the asymmetry of $\rho_s$ can be amplified under catalytic behavior with corresponding catalytic quantum states and global TIO channels. (b) Consider the mixed state of formula (\ref{state}), i.e., $x^2+z^2< 1$. Traverse the region with $x^2+z^2< 1, x\geq0, z\leq0$ and perform numerical calculations on the mixed states $\rho_s$ within it. The corresponding colors in the figure represent the increment of asymmetry of $\rho_s$, indicating that for almost all mixed states, the asymmetry can be amplified to a certain extent.}
\vspace{0.2cm}  
\label{figureS2}
\end{figure*}

 Our numerical result in the noiseless case is shown in Figure \ref{figureS2}. Here the dimension of the catalyst is limited to $2$. It is shown that even with the smallest catalyst, the asymmetry of almost all states in qubit system $\mathcal{S}$, i.e., states which are neither free states nor maximally resource states, can be amplified catalytically.
The detailed results in Figure \ref{figureS2}, indicate that, although the maximum increase in asymmetry of $\rho_s$ occurs when $|x|< |z|$, the asymmetry of $\rho_s$ can also be amplified to a certain extent when $|x|\approx |z|$ or $|x|> |z|$.

 \section{The effect of noise on catalytic amplification of asymmetry}
 Quantum catalysis is a highly sensitive task to noise. There are both TIO and non-TIO types of noise in the experiment, with non-TIO type noise having a more adverse impact on the catalytic task and even breaking constraint (\ref{yueshu}), making the task impossible to verify. In our experimental system, the corresponding non-TIO noise is mainly Pauli noise ($\sigma_x$). Next, based on this noise, we will explore the impact of noise on catalytic tasks.

 Let $\{|0\rangle,|1\rangle,|2\rangle,|3\rangle\}$ be the computational basis of the composed system of $\mathcal{S}$ and $\mathcal{C}$.
We believe that every two-dimensional subspace spanned by $\{|i\rangle,|j\rangle\}$ with $i,j=0,1,2,3$ and $i\neq j$, receives the same proportion of Pauli noise $\sigma_x^{ij}$. For the subspace spanned by $\{|0\rangle,|1\rangle\}$, the impact on the evolution process $\mathcal{E}(\cdot)=\sum_{i=1}^N K_i(\cdot)K_i^{\dagger}$ is:
\begin{align}
    \mathcal{E}^{01}(\cdot)=&p^{01}\sigma_x^{01}\mathcal{E}(\cdot)\sigma_x^{01}+(1-p^{01})\mathcal{E}(\cdot)\\
    =&p^{01}\sigma_x^{01}(\sum_{i=1}^N K_i(\cdot)K_i^{\dagger})\sigma_x^{01}+(1-p^{01})(\sum_{i=1}^N K_i(\cdot)K_i^{\dagger}) \label{noisy_model}
\end{align}
with
\begin{equation}
\begin{aligned}
&\sigma_x^{01}=\left[\begin{array}{cccc}
0 & 1 & 0 & 0 \\
1 & 0 & 0 & 0 \\
0 & 0 & 1 & 0 \\
0 & 0 & 0 & 1
\end{array}\right],
\end{aligned}
\end{equation}
where the superscript of $p^{01}$ represents the proportion of noise added to subspaces $|0\rangle$ and $|1\rangle$. Similarly, $p^{02}$, $p^{03}$, $p^{12}$, $p^{13}$ and $p^{23}$ are defined for the corresponding subspaces. Based on the channel $\mathcal{E}_{exp}$ in the experiment and the asymmetry introduced to systems $\mathcal{S}$ and $\mathcal{C}$ with $\epsilon_{\mathcal{S}}=0.0080$ and $\epsilon_{\mathcal{C}}=0.0073$, Our calculations show that $p^{ij}=0.003$ is quite consistent. Thus, experimental evaluation can be conducted on catalytic tasks for more quantum states, such as $|x|> |z|$ or $|x|\approx |z|$, which is different from the experimental protocol where $|x|< |z|$.

 \subsection{Case 1, where $|x|\approx|z|$}

 We provide a specific example and conduct an experimental evaluation on the current noise model of (\ref{noisy_model}) and noise ratio $p^{ij}=0.003$. The quantum state of system $\mathcal{S}$ is
\begin{align}
    \rho^1_\mathfrak{S}=\frac{1}{2}(\iden+0.7071\sigma_{x}-0.7071\sigma_{z}),
\end{align}
and the corresponding catalytic state $\rho^1_C$ reads
\begin{equation}
\rho^1_\mathfrak{C}=\frac{1}{2}\left(\mathbb{I}+0.6779 \sigma_{x}+0.3847\sigma_{z}\right).
\label{fuzhustate_case1}
\end{equation}
The Kraus operator corresponding to the global TIO $\mathcal{E}^1$ of this catalytic protocol is:
 
\begin{align}
\mathcal{K}_{0}^1&=\left[\begin{array}{cccc}
1 & 0 & 0 & 0 \\
0 & 0.0824 & 0.4435 & 0 \\
0 & -0.1772 & 0.8747 & 0 \\
0 & 0 & 0 & 0.7771
\end{array}\right],    
\label{KrausK0_case1}\\
\mathcal{K}_{1}^1&=\left[\begin{array}{cccc}
0 & \ \ \,0.9757 & 0.1358 & 0 \\
0 & 0 & 0 & 0.2944 \\
0 & 0 & 0 & 0.5416 \\
0 & 0 & 0 & 0
\end{array}\right],   
\label{KrausK1_case1}\\
\mathcal{K}_{2}^1&=\left[\begin{array}{cccc}
0 & \ -0.0995 & 0.1406 & 0 \\
0 & 0 & 0 & 0.0604 \\
0 & 0 & 0 & 0.1112 \\
0 & 0 & 0 & 0
\end{array}\right].   
\label{KrausK2_case1}
\end{align}

 In the above protocol, the ideal asymmetry increment of system $\mathcal{S}$ is 0.0811. Experimental noise (\ref{noisy_model}) is mixed into the ideal channel $\mathcal{E}^1$ to obtain the experimental simulated channel $\mathcal{E}_{simu}^1$ with process matrix $\mathcal{M}_{simu}^1$. Similar to Sec III, we modify the auxiliary state to study the asymmetry increment of system $\mathcal{S}$:
 
\begin{equation}
\rho^1_{\mathcal{C}}=\frac{1}{2}\left(\mathbb{I}+(0.6779+\delta x) \sigma_{x}+(0.3847+\delta z)\sigma_{z}\right) .
\label{fuzhustate_xiuzheng}
\end{equation}

\begin{figure*}[tph!]
\includegraphics [width= 0.6\textwidth]{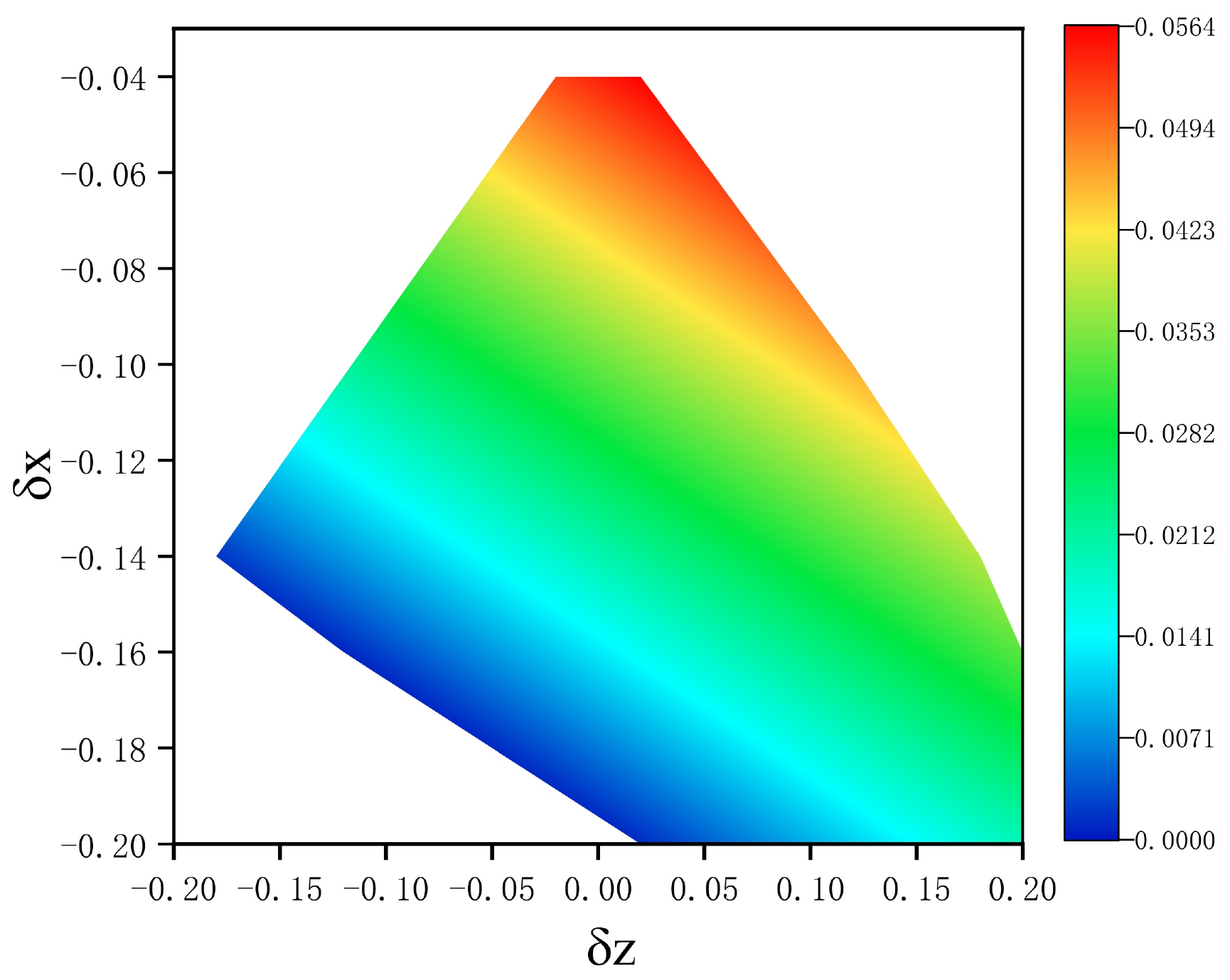}
\vspace{-0.3cm}%
\caption{ The range of catalytic states is optional under the current channel $\mathcal{M}_{simu}^1$ and the effective asymmetry increase of system $\mathcal{S}$. The auxiliary state in this region still satisfies constraint (\ref{yueshu}) after evolution with channel $\mathcal{E}_{simu}^1$.}
\vspace{0.2cm}  
\label{figureS3}
\end{figure*}

 As shown in Figure \ref{figureS3}, the asymmetry increase of system $\mathcal{S}$ under the current channel is $\Delta\eta_1=0.0564$ at most. This example looks better than the protocol in our experiment, but its evolution process is more complex, including three Kraus operators. However, under the current experimental noise, there is still a significant increase of the asymmetry of system $\mathcal{S}$ in this protocol, and the area available for auxiliary state selection is also relatively large, so it is very promising to implement it in experiments. Further simulation shows that when the noise ratio reaches $p^{ij}_{bound1}=0.010$, the region where the auxiliary state can be selected disappears, i.e. the catalytic protocol cannot proceed.

 \subsection{Case 2, where $|x|> |z|$}
 We provide a specific example and conduct an experimental evaluation on the current noise model of (\ref{noisy_model}) and noise ratio $p^{ij}=0.003$. The quantum state of system $\mathcal{S}$ is
\begin{align}
    \rho^2_\mathfrak{S}=\frac{1}{2}(\iden+0.8660\sigma_{x}-0.5000\sigma_{z}),
\end{align}
and the corresponding catalytic state $\rho^1_C$ reads
\begin{equation}
\rho^2_\mathfrak{C}=\frac{1}{2}\left(\mathbb{I}+0.7430 \sigma_{x}+0.4749\sigma_{z}\right) .\label{fuzhustate_case2}
\end{equation}
The Kraus operator corresponding to the global TIO $\mathcal{E}^2$ of this catalytic protocol is:
 
\begin{align}
\mathcal{K}_{0}^2&=\left[\begin{array}{cccc}
1 & 0 & 0 & 0 \\
0 & 0.1274 & 0.4352 & 0 \\
0 & -0.0603 & 0.8892 & 0 \\
0 & 0 & 0 & 0.8138
\end{array}\right],    \label{KrausK0_case2}\\
\mathcal{K}_{1}^2&=\left[\begin{array}{cccc}
0 & \ \ \,0.9864 & 0.0102 & 0 \\
0 & 0 & 0 & 0.2011 \\
0 & 0 & 0 & 0.5256 \\
0 & 0 & 0 & 0
\end{array}\right],   \label{KrausK1_case2}\\
\mathcal{K}_{2}^2&=\left[\begin{array}{cccc}
0 & \ -0.0841 & 0.1406 & 0 \\
0 & 0 & 0 & 0.0518 \\
0 & 0 & 0 & 0.1354 \\
0 & 0 & 0 & 0
\end{array}\right].   \label{KrausK2_case2}
\end{align}

 In the above protocol, the asymmetry increment of system $\mathcal{S}$ is 0.0405. Experimental noise (\ref{noisy_model}) is mixed into the ideal channel $\mathcal{E}^2$ to obtain the experimental simulated channel $\mathcal{E}_{simu}^2$ with process matrix $\mathcal{M}_{simu}^2$. Similar to Sec III, we modify the auxiliary state to study the asymmetry increment of system $\mathcal{S}$:
\begin{equation}
\rho^2_{\mathcal{C}}=\frac{1}{2}\left(\mathbb{I}+(0.7430+\delta x) \sigma_{x}+(0.4749+\delta z)\sigma_{z}\right) .
\label{fuzhustate_xiuzheng}
\end{equation}

\begin{figure*}[tph!]
\includegraphics [width= 0.6\textwidth]{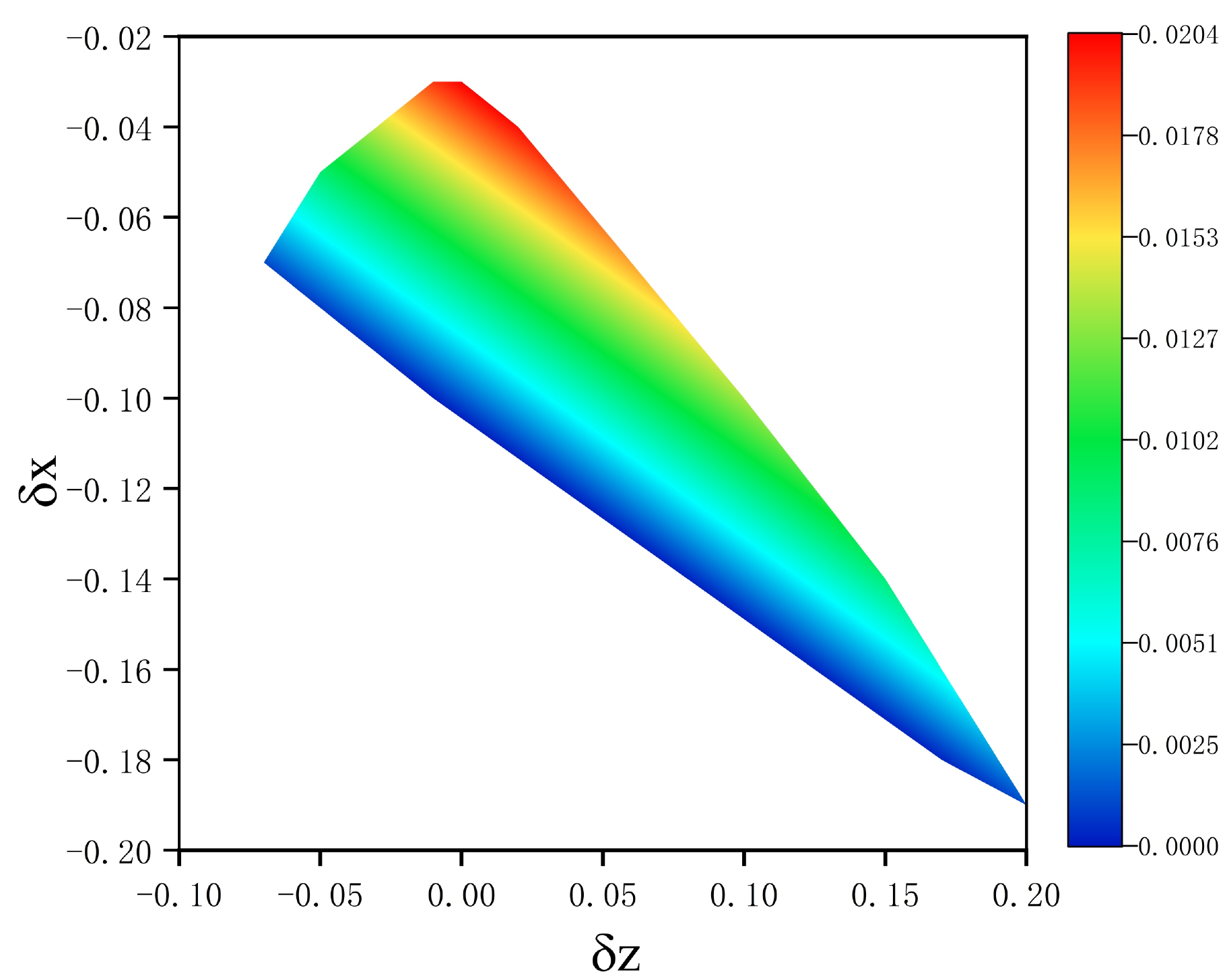}
\vspace{-0.3cm}%
\caption{ The range of catalytic states is optional under the current channel $\mathcal{M}_{simu}^2$ and the effective asymmetry increase of system $\mathcal{S}$. The auxiliary state in this region still satisfies constraint (\ref{yueshu}) after evolution with channel $\mathcal{E}_{simu}^2$.}
\vspace{0.2cm}  
\label{figureS4}
\end{figure*}

 As shown in Figure \ref{figureS4}, the asymmetry increase of system $\mathcal{S}$ under the current channel is $\Delta\eta_2=0.0204$ at most. Under the current experimental noise, there is a small increase of the asymmetry of system $\mathcal{S}$ in this protocol, and the area available for auxiliary state selection is also relatively small, so implementing it in experiments is relatively uncontrollable. Further simulation shows that when the noise ratio reaches $p^{ij}_{bound2}=0.006$, the region where the auxiliary state can be selected disappears, i.e., the catalytic protocol cannot proceed.

 We have provided a specific analysis of the two examples above, and of course, this process can be extended to any quantum state of system $\mathcal{S}$. Under constraint (\ref{yueshu}), the larger the proportion $p^{ij}$ of non-TIO noise in the channel, the fewer catalytic quantum states can be selected, and the smaller the amplification of the asymmetry of system $\mathcal{S}$ by the catalytic protocol. We also simulated the protocol used in the experiment, and when $p^{ij}_{bound}=0.014$, the catalytic protocol cannot proceed, indicating the sensitivity of the catalytic task to noise.


\newpage
\begin{table}[!htbp]
\caption{The real part of the process matrix $\mathbf{M}_p$}
     \centering
    \begin{tabular}{c|c|c|c|c|c|c|c|c|c|c|c|c|c|c|c}
    \hline
    0.9991 & 0.0015 & -0.0018 & 0.0005 & 0.0003 & -0.1563 & -0.3239 & -0.0012 & -0.0020 & 0.3968 & 0.8136 & -0.0006 & -0.0002 & 0.0030 & 0.0012 & 0.7416 \\ \hline
    0.0015 & 0.0003 & 0 & 0 & -0.0009 & -0.0003 & -0.0004 & 0.0002 & 0.0001 & 0.0015 & 0.0022 & -0.0001 & 0 & -0.0002 & -0.0006 & 0.0014\\ \hline
    -0.0018 & 0 & 0.0003 & 0 & 0.0004 & 0.0005 & 0 & 0 & 0.0003 & -0.0002 & 0.0005 & 0.0003 & -0.0001 & 0.0001 & 0.0001 & -0.0006\\ \hline
    0.0005 & 0 & 0 & 0.0003 & 0.0004 & 0 & -0.0002 & 0 & 0 & 0.0001 & 0.0003 & 0.0001 & 0.0001 & 0.0001 & -0.0004 & 0.0003\\ \hline
    0.0003 & -0.0009 & 0.0004 & 0.0004 & 0.8657 & 0.0058 & 0.0003 & 0.0008 & 0.3363 & -0.0136 & -0.0027 & 0.0010 & -0.0006 & 0.4340 & 0.4344 & -0.0043\\ \hline
    -0.1563 & -0.0003 & 0.0005 & 0 & 0.0058 & 0.0254 & 0.0508 & 0.0002 & 0.0024 & -0.0633 & -0.1275 & 0.0004 & -0.0002 & 0.0009 & 0.0017 & -0.1162\\ \hline
    -0.3239 & -0.0004 & 0 & -0.0002 & 0.0003 & 0.0508 & 0.1086 & 0.0001 & 0.0018 & -0.1300 & -0.2729 & -0.0003 & -0.0001 & -0.0008 & 0.0004 & -0.2437\\ \hline
    -0.0012 & 0.0002 & 0 & 0 & 0.0008 & 0.0002 & 0.0001 & 0.0003 & 0.0004 & -0.0004 & -0.0006 & -0.0001 & 0 & 0.0003 & -0.0002 & -0.0007\\ \hline
    -0.0020 & 0.0001 & 0.0003 & 0 & 0.3363 & 0.0024 & 0.0018 & 0.0004 & 0.1323 & -0.0040 & -0.0028 & 0.0003 & -0.0003 & 0.1698 & 0.1702 & -0.0033\\ \hline
    0.3968 & 0.0015 & -0.0002 & 0.0001 & -0.0136 & -0.0633 & -0.1300 & -0.0004 & -0.0040 & 0.1653 & 0.3348 & 0 & 0.0007 & -0.0021 & -0.0029 & 0.2981\\ \hline
    0.8136 & 0.0022 & 0.0005 & 0.0003 & -0.0027 & -0.1275 & -0.2729 & -0.0006 & -0.0028 & 0.3348 & 0.7021 & 0.0012 & 0.0011 & 0.0015 & 0.0023 & 0.6156\\ \hline
    -0.0006 & -0.0001 & 0.0003 & 0.0001 & 0.0010 & 0.0004 & -0.0003 & -0.0001 & 0.0003 & 0 & 0.0012 & 0.0004 & 0 & 0.0003 & 0.0006 & 0.0002\\ \hline
    -0.0002 & 0 & -0.0001 & 0.0001 & -0.0006 & -0.0002 & -0.0001 & 0 & -0.0003 & 0.0007 & 0.0011 & 0 & 0.0003 & 0.0001 & -0.0001 & 0.0001\\ \hline
    0.0030 & -0.0002 & 0.0001 & 0.0001 & 0.4340 & 0.0009 & -0.0008 & 0.0003 & 0.1698 & -0.0021 & 0.0015 & 0.0003 & 0.0001 & 0.2217 & 0.2203 & 0.0006\\ \hline
    0.0012 & -0.0006 & 0.0001 & -0.0004 & 0.4344 & 0.0017 & 0.0004 & -0.0002 & 0.1702 & -0.0029 & 0.0023 & 0.0006 & -0.0001 & 0.2203 & 0.2237 & -0.0009\\ \hline
    0.7416 & 0.0014 & -0.0006 & 0.0003 & -0.0043 & -0.1162 & -0.2437 & -0.0007 & -0.0033 & 0.2981 & 0.6156 & 0.0002 & 0.0001 & 0.0006 & -0.0009 & 0.5544\\ \hline
    \end{tabular} \label{real}
\end{table}

\begin{table}[!htbp]
\caption{The imaginary part of the process matrix $\mathbf{M}_p$}
     \centering
    \begin{tabular}{c|c|c|c|c|c|c|c|c|c|c|c|c|c|c|c}
    \hline
    0 & -0.0018 & 0.0019 & -0.0011 & 0.0001 & 0.0007 & -0.0007 & -0.0017 & -0.0021 & -0.0021 & -0.0053 & 0.0004 & -0.0001 & 0.0001 & -0.0014 & 0.0016)\\ \hline
    0.0018 & 0 & 0 & 0 & 0.0009 & -0.0002 & -0.0001 & -0.0002 & 0.0006 & 0.0013 & 0.0019 & 0.0001 & 0.0001 & 0.0005 & 0.0013 & 0.0012\\ \hline
    -0.0019 & 0 & 0 & 0 & -0.0021 & 0.0002 & 0.0001 &  0 & -0.0010 & -0.0003 & 0.0009 & 0 & 0.0001 & -0.0012 & -0.0009 & -0.0008\\ \hline
    0.0011 & 0 & 0 & 0 & 0 & 0.0003 & -0.0001 & 0 & -0.0002 & -0.0006 & 0.0001 & 0 & -0.0002 & -0.0010 & -0.0007 & 0.0004\\ \hline
    -0.0001 & -0.0009 & 0.0021 & 0 & 0 & 0.0016 & -0.0012 & 0 & -0.0001 & 0.0007 & 0.0001 & -0.0019 & -0.0018 & -0.0011 & -0.0032 & 0.0006\\ \hline
    -0.0007 & 0.0002 & -0.0002 & -0.0003 & -0.0016 & 0 & -0.0002 & 0.0002 & -0.0005 & -0.0002 & 0.0019 & 0 & -0.0002 & -0.0013 & 0.0001 & -0.0003\\ \hline
    0.0007 & 0.0001 & -0.0001 & 0.0001 & 0.0012 & 0.0002 & 0 & 0.0001 & 0.0015 & 0.0004 & 0.0004 & 0.0006 & -0.0005 & 0.0008 & 0.0019 & 0\\ \hline
    0.0017 & 0.0002 & 0 & 0 & 0 & -0.0002 & -0.0001 & 0 & 0.0005 & 0.0014 & 0.0019 & 0 & 0 & 0.0001 & 0.0004 & 0.0012\\ \hline
    0.0021 & -0.0006 & 0.0010 & 0.0002 & 0.0001 & 0.0005 & -0.0015 & -0.0005 & 0 & 0.0025 & 0.0048 & -0.0002 & -0.0005 & 0.0001 & 0 & 0.0027\\ \hline
    0.0021 & -0.0013 & 0.0003 & 0.0006 & -0.0007 & 0.0002 & -0.0004 & -0.0014 & -0.0025 & 0 & 0.0006 & 0.0004 & 0.0010 & -0.0008 & -0.0005 & 0.0017\\ \hline
    0.0053 & -0.0019 & -0.0009 & -0.0001 & -0.0001 & -0.0019 & -0.0004 & -0.0019 & -0.0048 & -0.0006 & 0 & -0.0014 & 0.0021 & -0.0010 & 0.0011 & 0.0026\\ \hline
    -0.0004 & -0.0001 & 0 & 0 & 0.0019 & 0 & -0.0006 & 0 & 0.0002 & -0.0004 & 0.0014 & 0 & 0.0001 & 0.0005 & 0.0007 & 0.0002\\ \hline
    0.0001 & -0.0001 & -0.0001 & 0.0002 & 0.0018 & 0.0002 & 0.0005 & 0 & 0.0005 & -0.0010 & -0.0021 & -0.0001 & 0 & 0.0004 & 0.0004 & -0.0007\\ \hline
    -0.0001 & -0.0005 & 0.0012 & 0.0010 & 0.0011 & 0.0013 & -0.0008 & -0.0001 & -0.0001 & 0.0008 & 0.0010 & -0.0005 & -0.0004 & 0 & -0.0030 & 0.0007\\ \hline
    0.0014 & -0.0013 & 0.0009 & 0.0007 & 0.0032 & -0.0001 & -0.0019 & -0.0004 & 0 & 0.0005 & -0.0011 & -0.0007 & -0.0004 & 0.0030 & 0 & 0.0012\\ \hline
    -0.0016 & -0.0012 & 0.0008 & -0.0004 & -0.0006 & 0.0003 & 0 & -0.0012 & -0.0027 & -0.0017 & -0.0026 & -0.0002 & 0.0007 & -0.0007 & -0.0012 & 0\\ \hline
    
    \end{tabular} \label{imag}
\end{table}

\bibliography{apssamp_revise}